\documentclass[12pt,preprint]{aastex}

\usepackage{graphicx}
\usepackage{epstopdf}
\usepackage{dblfloatfix}

\newcommand{\kms}{km s$^{-1}$}
\newcommand{\ha}{H$\alpha$}
\newcommand{\solar}{\ifmmode_{\sun}\else$_{\sun}$\fi}

\newcommand{\HI}{H$\,${\sc i}}

\newcommand{\rd}{$R_D$}
\newcommand{\rbr}{$R_{Br}$}

\newcommand{\coldens}{atoms cm$^{-2}$}

\begin{document}

\title{Gas engaged in noncircular motions in LITTLE THINGS dwarf irregular galaxies}

\author{
Deidre A.\ Hunter\altaffilmark{1},
Lauren Laufman\altaffilmark{1,2,3},
Se-Heon Oh\altaffilmark{4},
Stephen E.\ Levine\altaffilmark{1},
Caroline E.\ Simpson\altaffilmark{5}
}

\altaffiltext{1}{Lowell Observatory, 1400 West Mars Hill Road, Flagstaff, AZ 86001, USA}
\altaffiltext{2}{Department of Physics and Astronomy, Northern Arizona University, Flagstaff, AZ 86011, USA}
\altaffiltext{3}{Current address: Department of Astronomy, University of Wisconsin-Madison, 475 N. Charter St., Madison, WI 53706, USA}
\altaffiltext{4}{Department of Physics and Astronomy, Sejong University, 209 Neungdong-ro, Gwangjin-gu, Seoul, Republic of Korea}
\altaffiltext{5}{Department of Physics, Florida International University, CP 204, 11200 SW 8th St, Miami, FL 33199, USA}

\begin{abstract}
We have examined gas engaged in noncircular motions in 22 of the nearby LITTLE THINGS dwarf irregular galaxies.
The \HI\ data cubes have been deconvolved into kinematic components - bulk rotation and noncircular motions, to
produce maps of integrated gas, velocity field, and velocity dispersion in the different components.
We found significant regions of gas engaged in noncircular motions in half of the galaxies,
involving 1\%-20\% of the total \HI\ mass of the galaxy.
In one galaxy we found a pattern in the velocity field that is characteristic of 
streaming motions around the stellar bar potential and star formation at the end of bar.
%We see a similar velocity field pattern in DDO 47, but no stellar bar is obvious.
Two galaxies have large-scale filamentary structures found in their outer disks, and 
these filaments could be transient instabilities in the gas.
We found no spatial correlation between noncircular motion gas and enhanced star formation.
We found noncircular motion gas in only one galaxy associated with %an unusual section of the galaxy with
higher \HI\ velocity dispersion. % than elsewhere in that galaxy,
\end{abstract}


\section{Introduction} \label{sec-intro}

Most dwarf irregular galaxies (dIrrs) are engaged in on-going star formation at gas surface densities 
that are so low that, according to models, dIrrs should not be able to form 
star-forming clouds \citep{hunter98,barnes12,bigiel10,eh15}.
Thus, dIrrs are an extreme environment in which physical models for star formation and galaxy
growth can be observed and tested.

In some dIrrs gas is observed to be engaged in noncircular motions, and model dwarfs formed in cosmological simulations
also exhibit gas in noncircular motions due, for example, to strong azimuthal bisymmetric fluctuations \citep{read16,oman19}.
Collisions between noncirculary moving gas and gas engaged in
bulk rotation could provide one mechanism for bringing gas together into
clouds that are dense enough to be self-gravitating.
Star formation that is associated with noncircular motions in the atomic hydrogen \HI\ could be
triggered internally by, for example, flows around stellar bar potentials or stellar winds and explosions 
or triggered externally by interactions with another galaxy, ram pressure disturbances, or
extragalactic accretion. %, or other dynamical processes. 
Here we examine the nature of
noncircular motions of \HI\ and the role they play in the evolution of nearby dIrrs.
%We do this using the LITTLE THINGS \HI\ data cubes that have been deconvolved into circular and
%noncircular motions by \citet{oh15}.

\section{Data} \label{sec-data}

We used the LITTLE THINGS
(Local Irregulars That Trace Luminosity
Extremes, The \HI\ Nearby Galaxy Survey; Hunter et al.\ 2012) sample of dIrr galaxies.
LITTLE THINGS is
a multi-wavelength survey of 37 dIrr galaxies and 4 Blue Compact Dwarfs (BCDs) aimed at
understanding what drives star formation in tiny systems. The LITTLE THINGS galaxies
were chosen to be nearby ($\le$10.3 Mpc), contain gas so they could be forming stars, and cover
a large range in dwarf galactic properties, such as rate of star formation.
The LITTLE THINGS data sets include \HI\ spectral line maps obtained with the 
National Science Foundation's Karl G.\ Jansky Very Large Array
(VLA\footnote[6]{The VLA is a facility of the National Radio Astronomy Observatory.
The National Radio Astronomy Observatory is a facility of the National Science Foundation operated under cooperative agreement
by Associated Universities, Inc.}).
The \HI\ data cubes combine observations in the B, C, and D arrays and are characterized by
high sensitivity ($\le$1.1 mJy beam$^{-1}$ per channel), high spectral resolution (1.3 or 2.6 \kms), and
moderately high angular resolution (frequently, $\sim$6\arcsec).
Note that the pixel scale is 1.5\arcsec.
We also have ancillary multi-wavelength data including $UBV$, H$\alpha$, and {\it GALEX} FUV images.

This study includes the 22 LITTLE THINGS galaxies whose \HI\ rotation curves were analyzed by \citet{oh15}.
In an algorithm developed by \citet{oh08}, \citet{oh15} fit Gaussians to \HI\ position-velocity data cubes
at positions where multiple kinematic components were found.
Oh et al.\ used the \HI\ data cubes in which channel maps were made using robust weighting. This choice of weighting results in
maps with a
spatial resolution that is somewhat higher compared to that of naturally-weighted maps and the beam shape has less extended wings.
The kinematic components under consideration include ordered, circular rotation (``bulk'') and
noncircular motion.
They were deconvolved iteratively using the following steps:
1) estimate an initial rotation curve from ellipse fits to 3.6 $\mu$m images and single Gaussian fits to the intensity-velocity profiles
along the major axis of the galaxy in the \HI\ data cube, 
2) create a model velocity field,
3) perform single Gaussian fits across the data cube,
4) extract the bulk velocity field using multiple Gaussian decomposition,
and 5) iterate but now use a full tilted ring model to determine the rotation curve parameters.
In this way, the true bulk rotation could be determined and noncircular motions isolated.
Gas engaged in noncircular motions was classified as being strong noncircular (``snonc'')
if the intensity peak is higher than that of the bulk motion gas at that position 
or weak noncircular (``wnonc'') if the intensity peak is lower than that of the bulk motion gas at that position. 
We refer the reader to \citet{oh08,oh15} for a detailed description of the deconvolution process.

The final products of the deconvolution are separate maps of the bulk, snonc, and wnonc gas components.
\citet{oh15} made maps with signal-to-noise cuts at 2 and at 3. We chose to use the maps with a cut at a signal-to-noise of 3.
For each kinematic component, we have maps of the \HI\ integrated intensity (moment 0), velocity field (moment 1), and velocity dispersion (moment 2).
The galaxies used in this study and their observational and physical properties that are useful here are listed in Table \ref{tab-gal}.

As a check on the data, we summed, as an example, DDO 133's bulk0 and snonc0 maps (the wnonc0 map was not available) and compared
it to the total \HI\ mom0 map. We found that the sum of the components reproduces the total \HI\ map 
both in terms of internal morphology and integrated intensity but with increased noise.
In some of the galaxy bulk maps, the places where large snonc regions occur appear as blank. 
This is due to the signal-to-noise cut that eliminates the low signal-to-noise bulk motion gas where it is dominated by
the snonc. We can see that all kinematic components are, nevertheless, present in those regions from intensity-velocity plots 
at spots within the snonc regions shown below in Section 3.2.

\begin{deluxetable}{lcccccccccc}
%\tabletypesize{\scriptsize}
\tabletypesize{\tiny}
\rotate
\tablecaption{Sample Galaxies\label{tab-gal}}
\tablewidth{0pt}
\tablehead{
\colhead{} & \colhead{D} & \colhead{Beam size\tablenotemark{a}}
& \colhead{Channel width}
& \colhead{$\log \rm SFR_D^{FUV}$\tablenotemark{b}}
& \colhead{$\log M_{\rm HI~tot}$\tablenotemark{c}}
& \colhead{$\log M_{\rm stars}$\tablenotemark{d}}
& \colhead{$R_{\rm Br}$\tablenotemark{e}}
& \colhead{$R_D$\tablenotemark{e}}
& \colhead{} 
& \colhead{} \\
\colhead{Galaxy} & \colhead{(Mpc)} & \colhead{(pc)}
& \colhead{(\kms)}
& \colhead{($M$\solar yr$^{-1}$ kpc$^{-2}$)}
& \colhead{($M$\solar)}  & \colhead{($M$\solar)}  & \colhead{(kpc)}
& \colhead{(kpc)} 
& \colhead{$N_{reg}$\tablenotemark{f}}
& \colhead{$M_{\rm snonc}/M_{\rm HI tot}$\tablenotemark{g}} 
}
\startdata
CVnIdwA  &  3.6   &  187 & 1.29 & -2.48$\pm$0.01 & 7.67 & 6.69     &  0.56$\pm$0.49 &  0.25$\pm$0.12 &   2 &  0.13  \\
DDO 43      &  7.8 &  262 & 2.58 & -1.55$\pm$0.01 & 8.23 & \nodata & 1.46$\pm$0.53 &  0.87$\pm$0.10 &   0 &  0      \\
DDO 46      &  6.1 &  170 & 2.58 & -2.46$\pm$0.01 & 8.27 & \nodata & 1.27$\pm$0.18 &  1.13$\pm$0.05 &   0 &  0      \\
DDO 47      &  5.2 &  245 & 2.58 & -2.40$\pm$0.01 & 8.59 & \nodata & \nodata              &  1.34$\pm$0.05 & 12 &  0.07 \\
DDO 50      &  3.4 &  108 & 2.58 & -1.55$\pm$0.01 & 8.85 & 8.03      & 2.65$\pm$0.27 &  1.48$\pm$0.06 &   4 &  0.02  \\
DDO 52      & 10.3 &  297 & 2.58 & -2.43$\pm$0.01 & 8.43 & 7.73     &  2.80$\pm$1.35 &  1.26$\pm$0.04 &   0 & 0       \\
DDO 53      &  3.6  &  105 & 2.58 & -2.41$\pm$0.01 & 7.72 & 6.99     &  0.62$\pm$0.09 &  0.47$\pm$0.01 &   1 & 0.01  \\
DDO 70      &  1.3  &    85 & 1.29 & -2.16$\pm$0.00 & 7.61 & 7.29     &  0.13$\pm$0.07 &  0.44$\pm$0.01 &  14 & 0.04   \\
DDO 87       &  7.7 &  256 & 2.58 & -1.00$\pm$0.01 & 8.39 & 7.51    &  0.99$\pm$0.11 &  1.21$\pm$0.02 &    0 & 0       \\
DDO 101     &  6.4 &  236 & 2.59 & -2.81$\pm$0.01 & 7.36 & 7.82    & 1.16$\pm$0.11 &  0.97$\pm$0.06 &    0 & 0       \\
DDO 126     &  4.9 &  147 & 2.58 & -2.10 $\pm$0.01 & 8.16 & 7.21   & 0.60$\pm$0.05 &  0.84$\pm$0.13 &    6 & 0.08 \\
DDO 133     &  3.5 &  195 & 2.58 & -2.62$\pm$0.01 & 8.02 & 7.48    &  2.25$\pm$0.24 &  1.22$\pm$0.04 &  18  & 0.11  \\
DDO 154     &  3.7 &   127 & 2.58 & -1.93$\pm$0.01 & 8.46 & 6.92   &  0.62$\pm$0.09 &  0.48$\pm$0.02 &   0  & 0      \\
DDO 168     &  4.3 &   141 & 2.58 & -2.04$\pm$0.01 & 8.45 & 7.77   &  0.72$\pm$0.07 &  0.83$\pm$0.01 &   5  & 0.04  \\
DDO 210     &  0.9 &     44 & 1.29 & -2.71$\pm$0.06 & 6.30 & 5.78   & \nodata &  0.16$\pm$0.01              &   4  & 0.20 \\
F564-V3      &  8.7 &    424 & 1.29 & -2.79$\pm$0.02 & 7.61 & \nodata &  0.73$\pm$0.40 &  0.63$\pm$0.09 &  0  & 0     \\
IC 1613       &  0.7 &      24 & 2.57 & -1.99$\pm$0.01 & 7.53 & 7.46      &  0.71$\pm$0.12 &  0.53$\pm$0.02 & 18  & 0.04 \\
NGC 3738    &  4.9 &    140 & 2.58 & -1.53$\pm$0.01 & 8.06 & 8.67    &  1.16$\pm$0.20 &  0.77$\pm$0.01 &  0  & 0     \\
UGC 8508    &  2.6 &      68 & 1.29 & \nodata               & 7.28 & 6.89   &  0.41$\pm$0.06 &  0.23$\pm$0.0   &  0  & 0     \\
WLM             &  1.0 &      30 & 2.58 & -2.05$\pm$0.01 & 7.85 & 7.21   &  0.83$\pm$0.16 &  1.18$\pm$0.24 &  0 & 0      \\
Haro 29         &  5.8 &    173 & 2.58 & -1.07$\pm$0.01 & 7.80 & 7.16   &  1.15$\pm$0.26 &  0.33$\pm$0.00 &   0 & 0     \\
Haro 36         &  9.3 &    286 & 2.58 & -1.55$\pm$0.01 & 8.16 & \nodata & 1.16$\pm$0.13 &  1.01$\pm$0.00 &   5 & 0.14 \\
\enddata
\tablenotetext{a}{VLA beam FWHM in parsecs determined from the square-root of the major axis times the minor axis, for
the robust-weighted maps \citep{lt12}.}
\tablenotetext{b}{Integrated star formation rate (SFR) determined from the {\it GALEX} FUV luminosity \citep{ludka}. The SFR is 
normalized by dividing by the area within one disk scale length \rd.}
\tablenotetext{c}{Total \HI\ mass of the galaxy from \citet{lt12}.}
\tablenotetext{d}{Total stellar mass of the galaxy determined from spectral energy distribution fitting from \citet{oh15}.
Galaxies without stellar mass values do not have stellar masses given by \citet{oh15} because the
full complement of passbands was not available for those galaxies \citep{zhang12}. 
All but two of the galaxies listed here have \HI\ masses greater than the stellar mass, and
the median \HI\ to stellar mass ratio is 4.4.}
\tablenotetext{e}{\rbr\ is the radius at which the $V$-band surface brightness profile changes slope, if it does. \rd\ is the disk scale length
 determined from the $V$-band \citep{breaks}.}
 \tablenotetext{f}{Number of strong noncircular motion regions analyzed here.}
\tablenotetext{g}{Integrated \HI\ mass in noncircular motion divided by total \HI\ gas mass.
Galaxies with no snonc regions analyzed here ($N_{reg}$=0) have $M_{\rm snonc}$ of 0.}
\end{deluxetable}

\begin{figure}[t!]
\vskip -0.5truein
\includegraphics[scale=0.825]{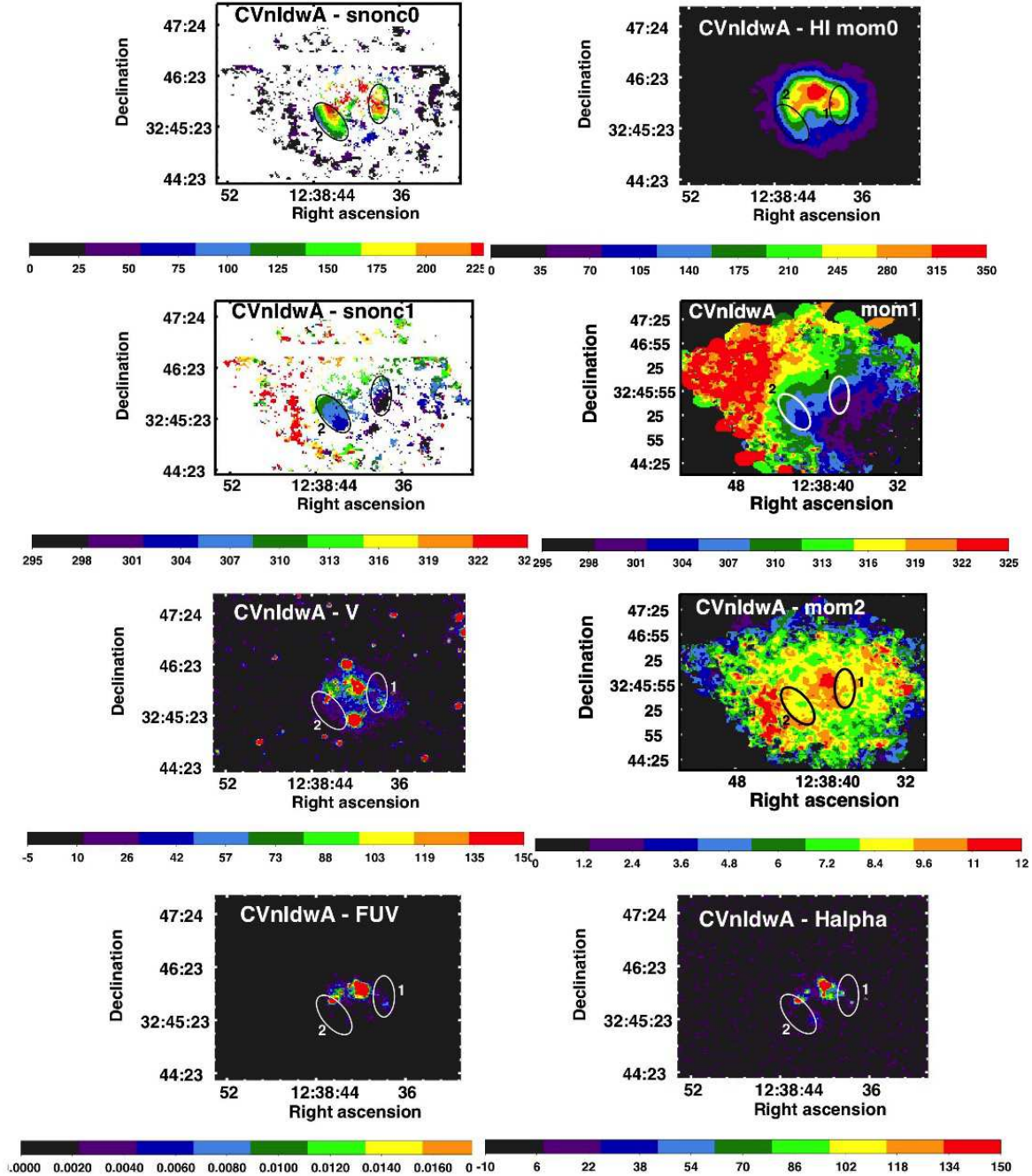}
%\vskip 0.5truein
\caption{For CVnIdwA, features identified on strong noncircular moment 0 (snonc0) maps
shown on snonc0, strong non-circular moment 1 (``snonc1''), total \HI\ moment 0 (``mom0''), total \HI\ moment 1 (``mom1''), total \HI\ moment 2 (``mom2''),
FUV, \ha\, and $V$-band maps.
Properties of the regions are given in Table \ref{tab-reg}.
Colorbar units are as follows: moment 0 maps are Jy beam$^{-1}$ m s$^{-1}$; 
snonc1, total \HI\ mom1, and total \HI\ mom2 maps are km s$^{-1}$;
FUV image is counts s$^{-1}$ that can be converted to erg s$^{-1}$ cm$^{-2}$ \AA$^{-1}$ by multiplying by $1.4\times10^{-15}$;
\ha\ image is counts that can be converted to erg s$^{-1}$ cm$^{-2}$ by multiplying by $3.975\times10^{-18}$  \citep{he04};
and $V$-band image is counts that can be converted to a Johnson $V$ magnitude using 
%$-2.5\log({\rm counts}/600) = V - 25 + 3.244 + 0.152\times1.27 - 0.013\times(B-V)$ \citep{he06}.
$-2.5\log({\rm counts}/600) = V - 21.56 - 0.013\times(B-V)$ \citep{he06}.
\label{fig-maps1}}
\end{figure}

\begin{figure}
\vskip -0.85truein
\includegraphics[scale=0.825]{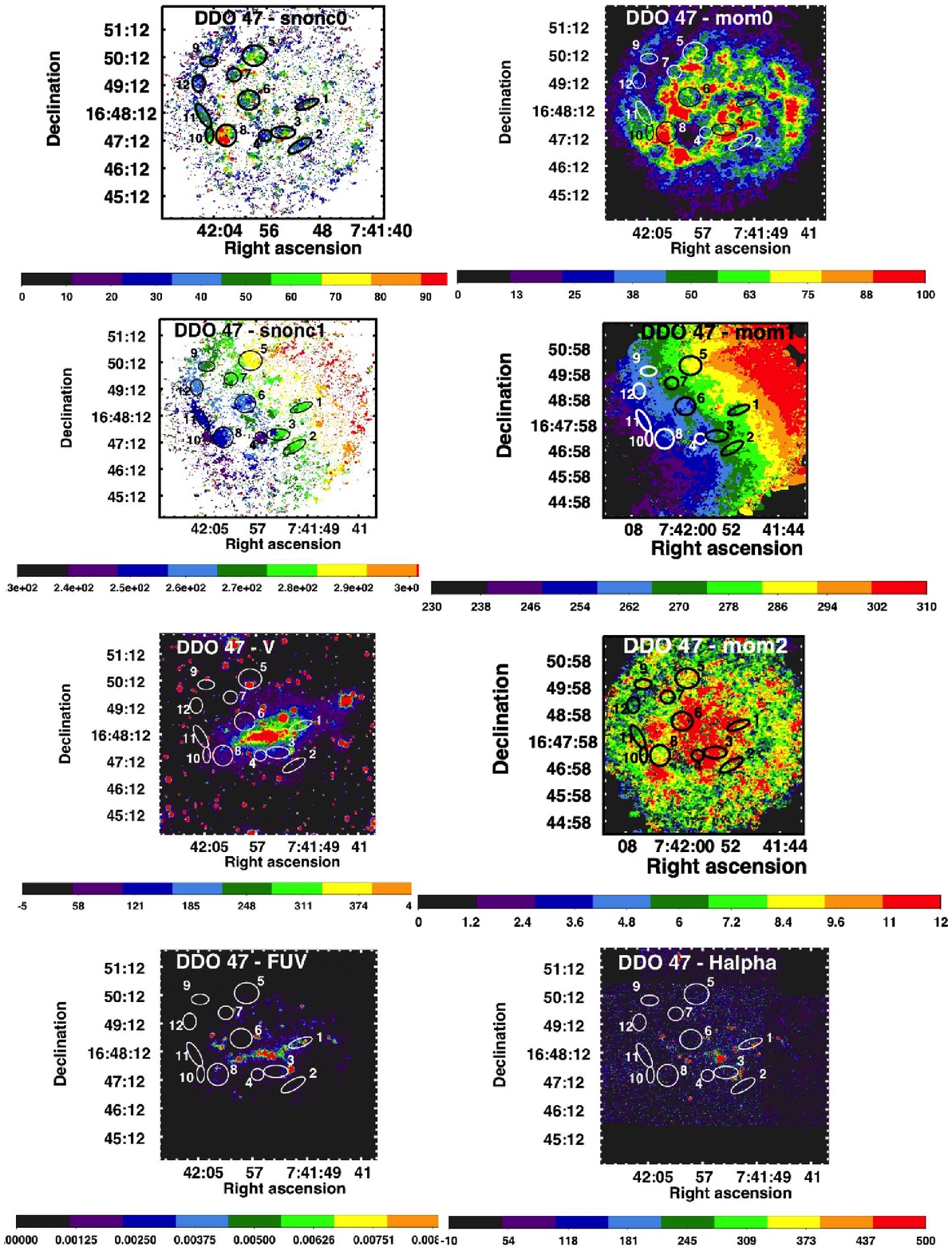}
%\vskip -0.8truein
\caption{For DDO 47, features identified on strong noncircular moment 0 (snonc0) maps
shown on snonc0, strong non-circular moment 1 (``snonc1''), total \HI\ moment 0 (``mom0''), total \HI\ moment 1 (``mom1''), total \HI\ moment 2 (``mom2''),
FUV, \ha\, and $V$-band maps.
Properties of the regions are given in Table \ref{tab-reg}.
Colorbar units are as in Figure \ref{fig-maps1}.
%follows: moment 0 maps are Jy beam$^{-1}$ m s$^{-1}$; snonc1 is km s$^{-1}$;
%total \HI\ moment 1 and moment 2 maps are m s$^{-1}$; FUV image is counts s$^{-1}$ 
%that can be converted to erg s$^{-1}$ cm$^{-2}$ \AA$^{-1}$ by multiplying by $1.4\times10^{-15}$;
\ha\ image is counts that can be converted to erg s$^{-1}$ cm$^{-2}$ by multiplying by $0.356\times10^{-18}$ \citep{he04};
and $V$-band image is counts that can be converted to a Johnson $V$ magnitude using 
%$-2.5\log({\rm counts}/1200) = V - 25 + 3.017 + 0.129\times1.17 - 0.045\times(B-V)$ \citep{he06}.
$-2.5\log({\rm counts}/1200) = V -21.83 - 0.045\times(B-V)$ \citep{he06}.
\label{fig-maps2}}
\end{figure}

\begin{figure}
\vskip -0.85truein
\includegraphics[scale=0.825]{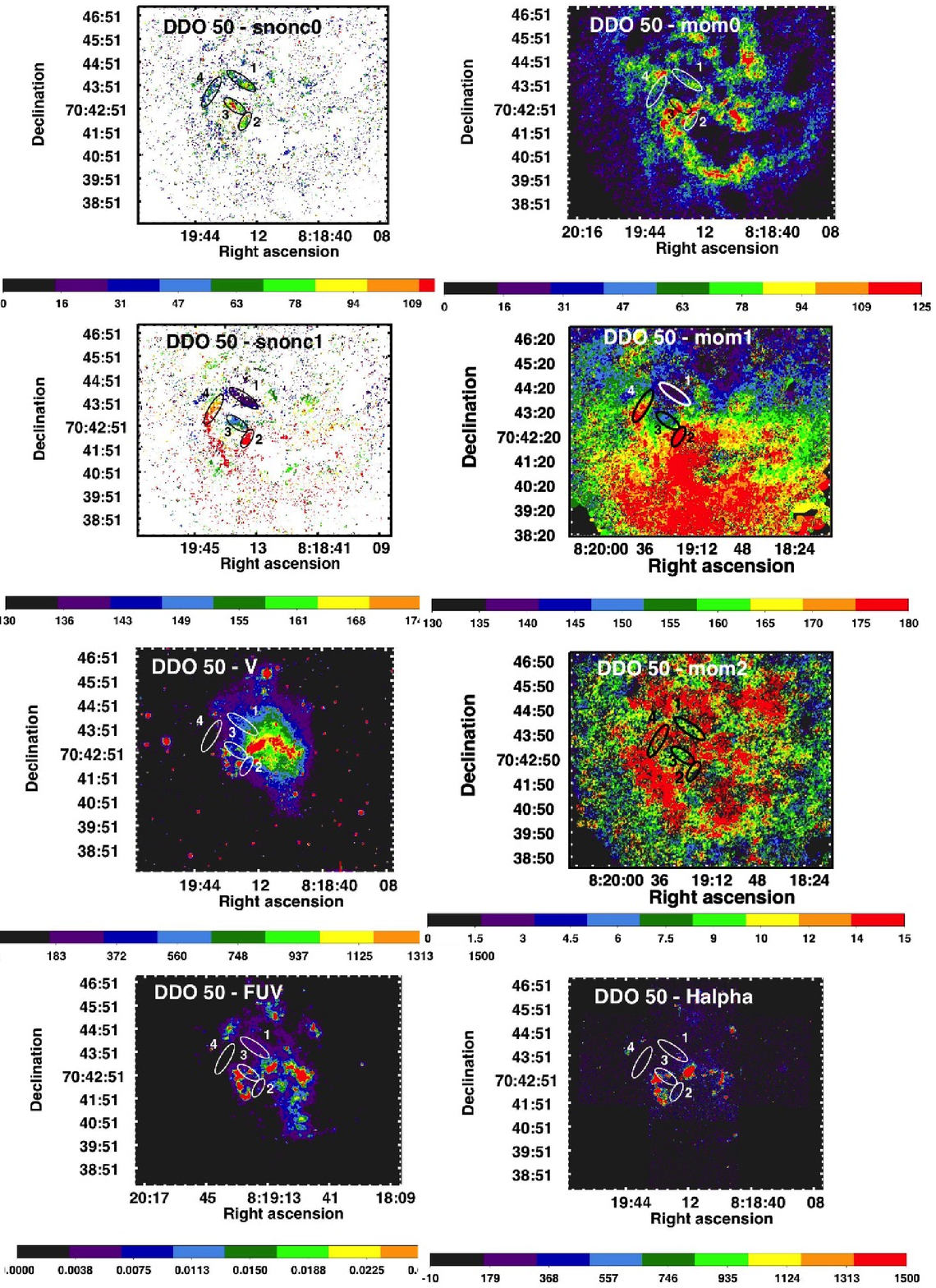}
%\vskip -0.9truein
\caption{For DDO 50, features identified on strong noncircular moment 0 (snonc0) maps
shown on snonc0, strong non-circular moment 1 (``snonc1''), total \HI\ moment 0 (``mom0''), total \HI\ moment 1 (``mom1''), total \HI\ moment 2 (``mom2''),
FUV, \ha\, and $V$-band maps.
Properties of the regions are given in Table \ref{tab-reg}.
Colorbar units are as in Figure \ref{fig-maps1}.
%Colorbar units are as follows: moment 0 maps are Jy beam$^{-1}$ m s$^{-1}$; snonc1 is km s$^{-1}$;
%total \HI\ moment 1 and moment 2 maps are m s$^{-1}$; FUV image is counts s$^{-1}$ 
%that can be converted to erg s$^{-1}$ cm$^{-2}$ \AA$^{-1}$ by multiplying by $1.4\times10^{-15}$;
\ha\ image is counts that can be converted to erg s$^{-1}$ cm$^{-2}$ by multiplying by $0.667\times10^{-18}$ \citep{he04};
and $V$-band image is counts that can be converted to a Johnson $V$ magnitude using 
%$-2.5\log({\rm counts}/1200) = V - 25 + 3.122 + 0.129\times1.25 - 0.016\times(B-V)$ \citep{he06}.
$-2.5\log({\rm counts}/1200) = V - 21.72 - 0.016\times(B-V)$ \citep{he06}.
\label{fig-maps3}}
\end{figure}

\begin{figure}
\vskip -0.85truein
\includegraphics[scale=0.825]{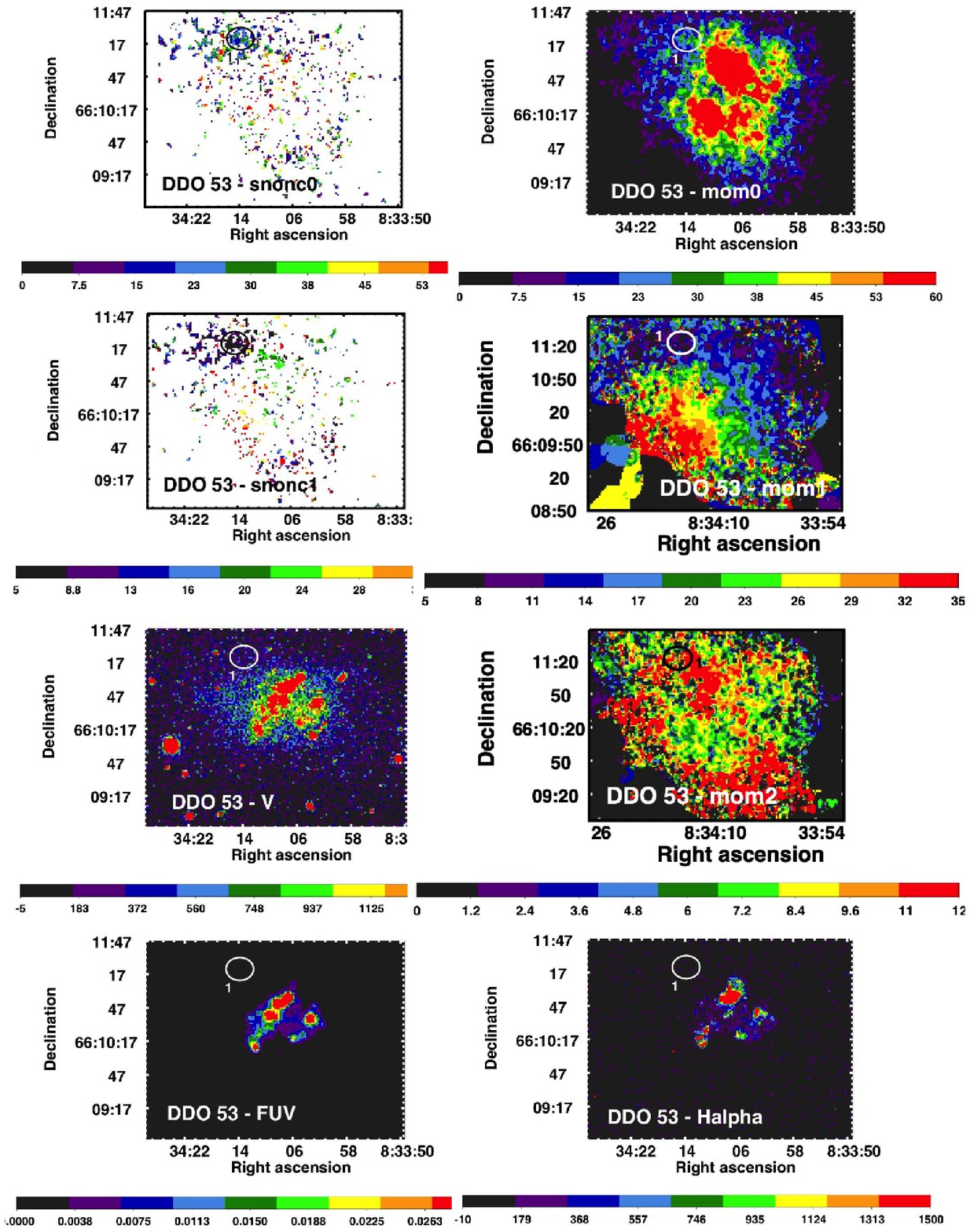}
%\vskip -1.0truein
\caption{For DDO 53, features identified on strong noncircular moment 0 (snonc0) maps
shown on snonc0, strong non-circular moment 1 (``snonc1''), total \HI\ moment 0 (``mom0''), total \HI\ moment 1 (``mom1''), total \HI\ moment 2 (``mom2''),
FUV, \ha\, and $V$-band maps.
Properties of the regions are given in Table \ref{tab-reg}.
Colorbar units are as in Figure \ref{fig-maps1}.
%Colorbar units are as follows: moment 0 maps are Jy beam$^{-1}$ m s$^{-1}$; snonc1 is km s$^{-1}$;
%total \HI\ moment 1 and moment 2 maps are m s$^{-1}$; FUV image is counts s$^{-1}$ 
%that can be converted to erg s$^{-1}$ cm$^{-2}$ \AA$^{-1}$ by multiplying by $1.4\times10^{-15}$;
\ha\ image is counts that can be converted to erg s$^{-1}$ cm$^{-2}$ by multiplying by $0.479\times10^{-18}$ \citep{he04};
and $V$-band image is counts that can be converted to a Johnson $V$ magnitude using 
%$-2.5\log({\rm counts}/180) = V - 25 - 0.297 + 0.2\times1.3 - 0.032\times(B-V)$ \citep{he06}.
$-2.5\log({\rm counts}/180) = V - 25.04 - 0.032\times(B-V)$ \citep{he06}.
\label{fig-maps4}}
\end{figure}

\begin{figure}
\vskip -0.95truein
\includegraphics[scale=0.825]{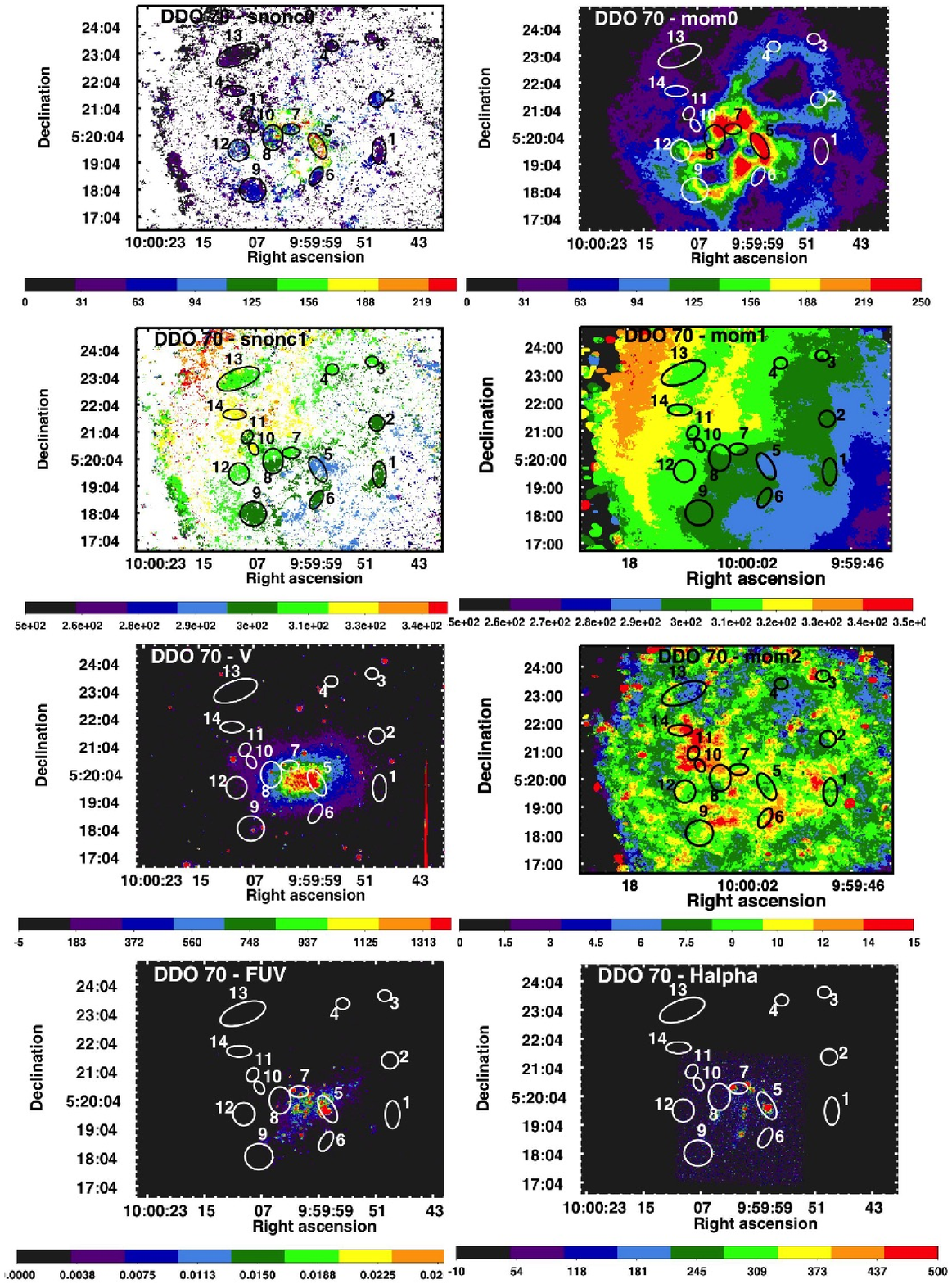}
%\vskip -0.7truein
\caption{For DDO 70, features identified on strong noncircular moment 0 (snonc0) maps
shown on snonc0, strong non-circular moment 1 (``snonc1''), total \HI\ moment 0 (``mom0''), total \HI\ moment 1 (``mom1''), total \HI\ moment 2 (``mom2''),
FUV, \ha\, and $V$-band maps.
Properties of the regions are given in Table \ref{tab-reg}.
Colorbar units are as in Figure \ref{fig-maps1}.
%Colorbar units are as follows: moment 0 maps are Jy beam$^{-1}$ m s$^{-1}$; snonc1 is km s$^{-1}$;
%total \HI\ moment 1 and moment 2 maps are m s$^{-1}$; FUV image is counts s$^{-1}$ 
%that can be converted to erg s$^{-1}$ cm$^{-2}$ \AA$^{-1}$ by multiplying by $1.4\times10^{-15}$;
\ha\ image is counts that can be converted to erg s$^{-1}$ cm$^{-2}$ by multiplying by $0.673\times10^{-18}$ \citep{he04};
and $V$-band image is counts that can be converted to a Johnson $V$ magnitude using 
%$-2.5\log({\rm counts}/1200) = V - 25 + 3.273 + 0.017\times(B-V)$ \citep{he06}.
$-2.5\log({\rm counts}/1200) = V -21.73 + 0.017\times(B-V)$ \citep{he06}.
\label{fig-maps5}}
\end{figure}

\begin{figure}
\vskip -0.95truein
\includegraphics[scale=0.825]{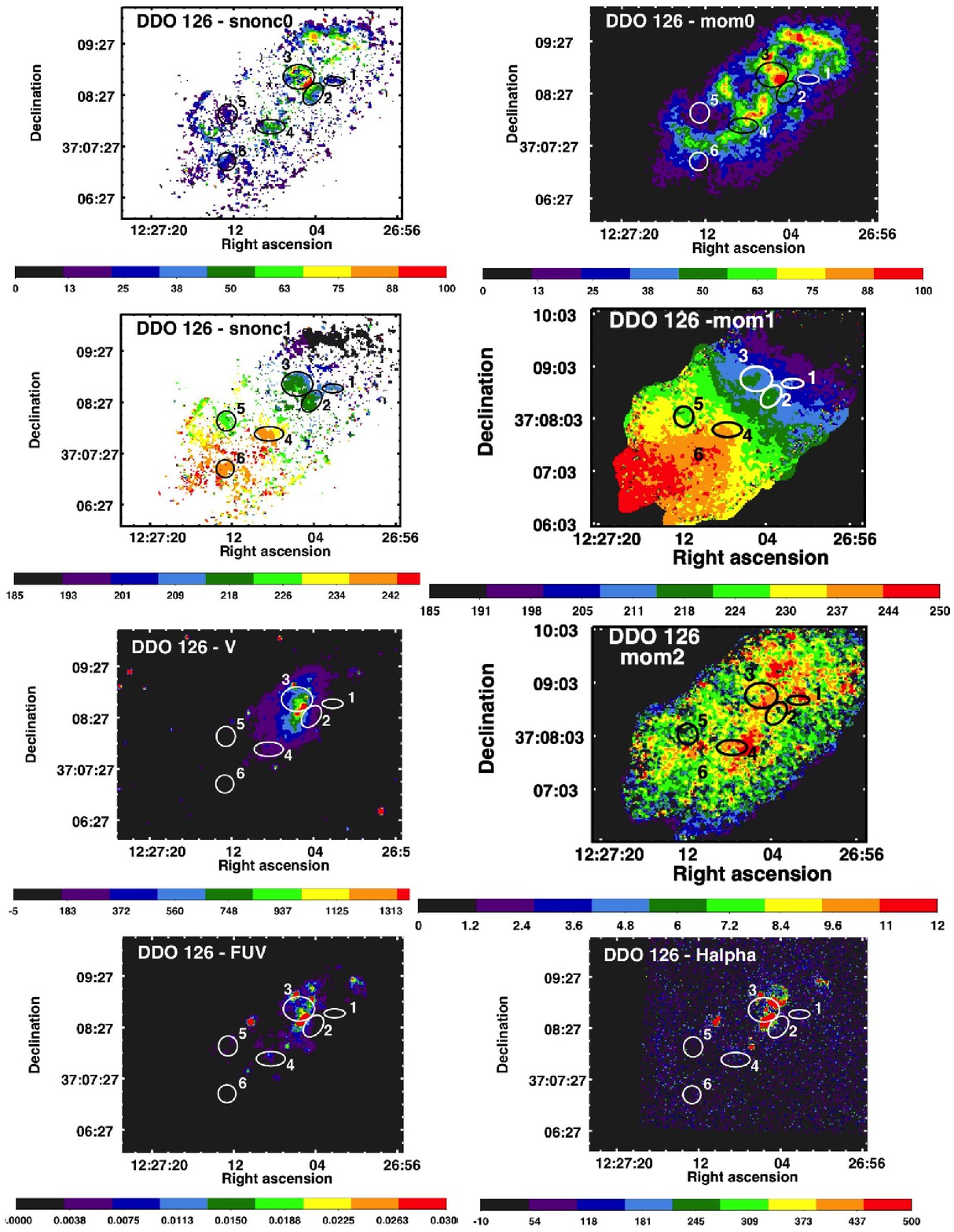}
%\vskip -0.8truein
\caption{For DDO 126, features identified on strong noncircular moment 0 (snonc0) maps
shown on snonc0, strong non-circular moment 1 (``snonc1''), total \HI\ moment 0 (``mom0''), total \HI\ moment 1 (``mom1''), total \HI\ moment 2 (``mom2''),
FUV, \ha\, and $V$-band maps.
Properties of the regions are given in Table \ref{tab-reg}.
Colorbar units are as in Figure \ref{fig-maps1}.
%Colorbar units are as follows: moment 0 maps are Jy beam$^{-1}$ m s$^{-1}$; snonc1 is km s$^{-1}$;
%total \HI\ moment 1 and moment 2 maps are m s$^{-1}$; FUV image is counts s$^{-1}$ 
%that can be converted to erg s$^{-1}$ cm$^{-2}$ \AA$^{-1}$ by multiplying by $1.4\times10^{-15}$;
\ha\ image is counts that can be converted to erg s$^{-1}$ cm$^{-2}$ by multiplying by $0.678\times10^{-18}$ \citep{he04};
and $V$-band image is counts that can be converted to a Johnson $V$ magnitude using 
%$-2.5\log({\rm counts}/1200) = V - 25 + 3.246 + 0.115\times1.04 - 0.017\times(B-V)$ \citep{he06}.
$-2.5\log({\rm counts}/1200) = V - 21.63 - 0.017\times(B-V)$ \citep{he06}.
\label{fig-maps6}}
\end{figure}

\begin{figure}
\vskip -0.95truein
\includegraphics[scale=0.825]{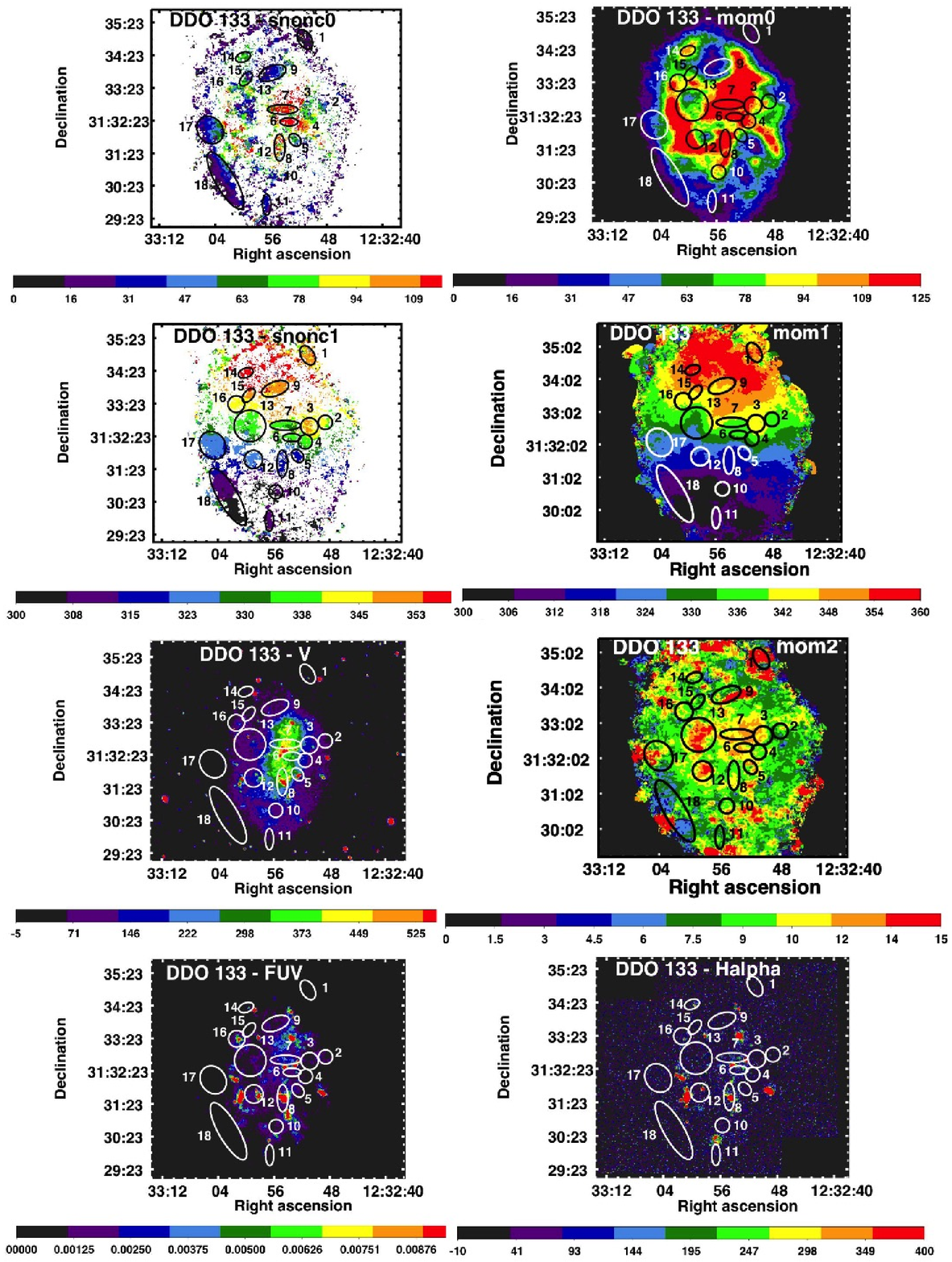}
%\vskip -1.0truein
\caption{For DDO 133, features identified on strong noncircular moment 0 (snonc0) maps
shown on snonc0, strong non-circular moment 1 (``snonc1''), total \HI\ moment 0 (``mom0''), total \HI\ moment 1 (``mom1''), total \HI\ moment 2 (``mom2''),
FUV, \ha\, and $V$-band maps.
Properties of the regions are given in Table \ref{tab-reg}.
Colorbar units are as in Figure \ref{fig-maps1}.
%Colorbar units are as follows: moment 0 maps are Jy beam$^{-1}$ m s$^{-1}$; snonc1 is km s$^{-1}$;
%total \HI\ moment 1 and moment 2 maps are m s$^{-1}$; FUV image is counts s$^{-1}$ 
%that can be converted to erg s$^{-1}$ cm$^{-2}$ \AA$^{-1}$ by multiplying by $1.4\times10^{-15}$;
\ha\ image is counts that can be converted to erg s$^{-1}$ cm$^{-2}$ by multiplying by $0.419\times10^{-18}$ \citep{he04};
and $V$-band image is counts that can be converted to a Johnson $V$ magnitude using 
%$-2.5\log({\rm counts}/1200) = V - 25 + 3.246 + 0.101\times1.035 - 0.017\times(B-V)$ \citep{he06}.
$-2.5\log({\rm counts}/1200) = V - 21.65 - 0.017\times(B-V)$ \citep{he06}.
\label{fig-maps7}}
\end{figure}

\begin{figure}
\vskip -0.95truein
\includegraphics[scale=0.825]{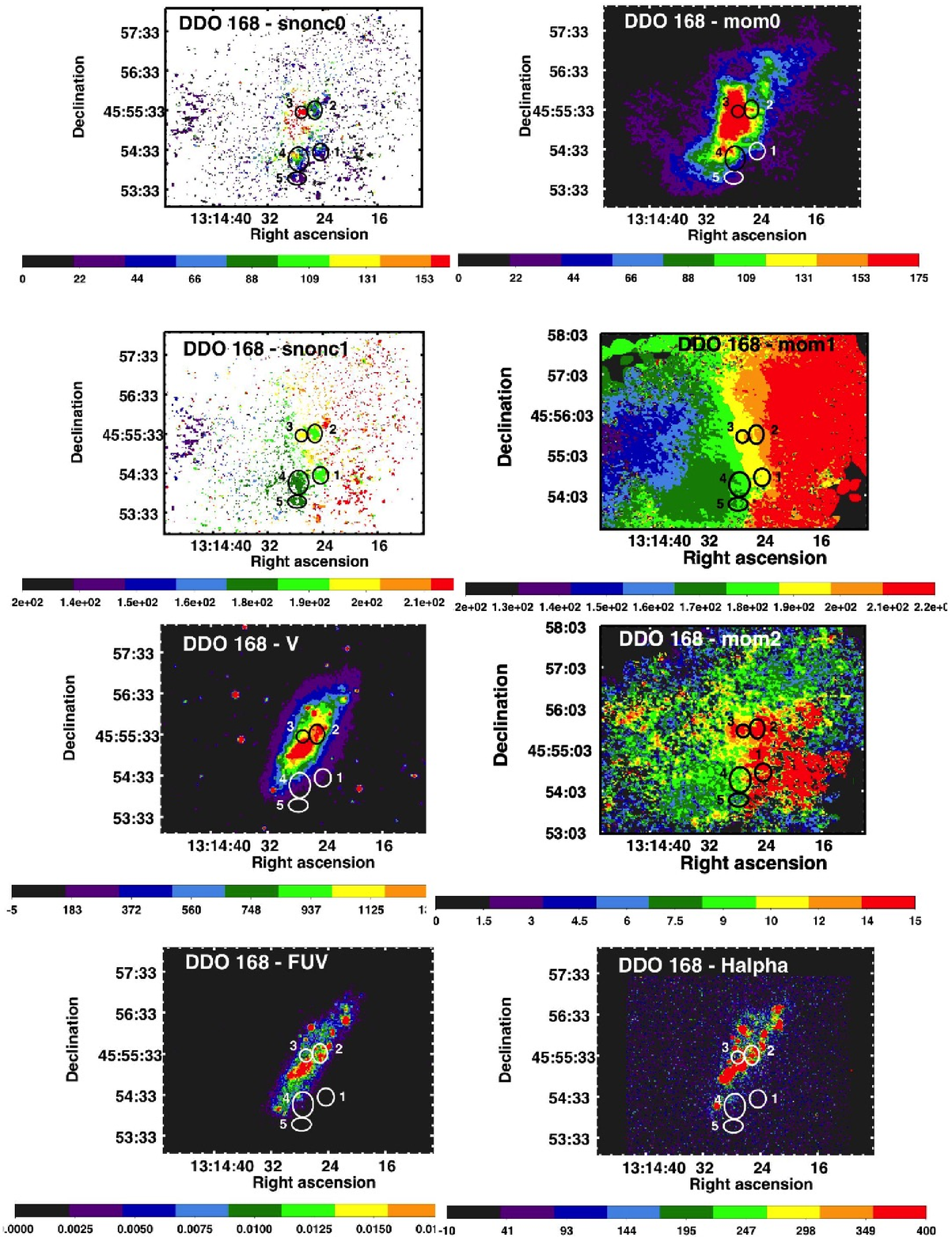}
%\vskip -1.0truein
\caption{For DDO 168, features identified on strong noncircular moment 0 (snonc0) maps
shown on snonc0, strong non-circular moment 1 (``snonc1''), total \HI\ moment 0 (``mom0''), total \HI\ moment 1 (``mom1''), total \HI\ moment 2 (``mom2''),
FUV, \ha\, and $V$-band maps.
Properties of the regions are given in Table \ref{tab-reg}.
Colorbar units are as in Figure \ref{fig-maps1}.
%Colorbar units are as follows: moment 0 maps are Jy beam$^{-1}$ m s$^{-1}$; snonc1 is km s$^{-1}$;
%total \HI\ moment 1 and moment 2 maps are m s$^{-1}$; FUV image is counts s$^{-1}$ 
%that can be converted to erg s$^{-1}$ cm$^{-2}$ \AA$^{-1}$ by multiplying by $1.4\times10^{-15}$;
\ha\ image is counts that can be converted to erg s$^{-1}$ cm$^{-2}$ by multiplying by $0.435\times10^{-18}$ \citep{he04};
and $V$-band image is counts that can be converted to a Johnson $V$ magnitude using 
%$-2.5\log({\rm counts}/1200) = V - 25 + 3.017 + 0.151\times1.08 + 0.045\times(B-V)$ \citep{he06}.
$-2.5\log({\rm counts}/1200) = V - 21.82 + 0.045\times(B-V)$ \citep{he06}.
\label{fig-maps8}}
\end{figure}

\begin{figure}
\vskip -0.95truein
\includegraphics[scale=0.825]{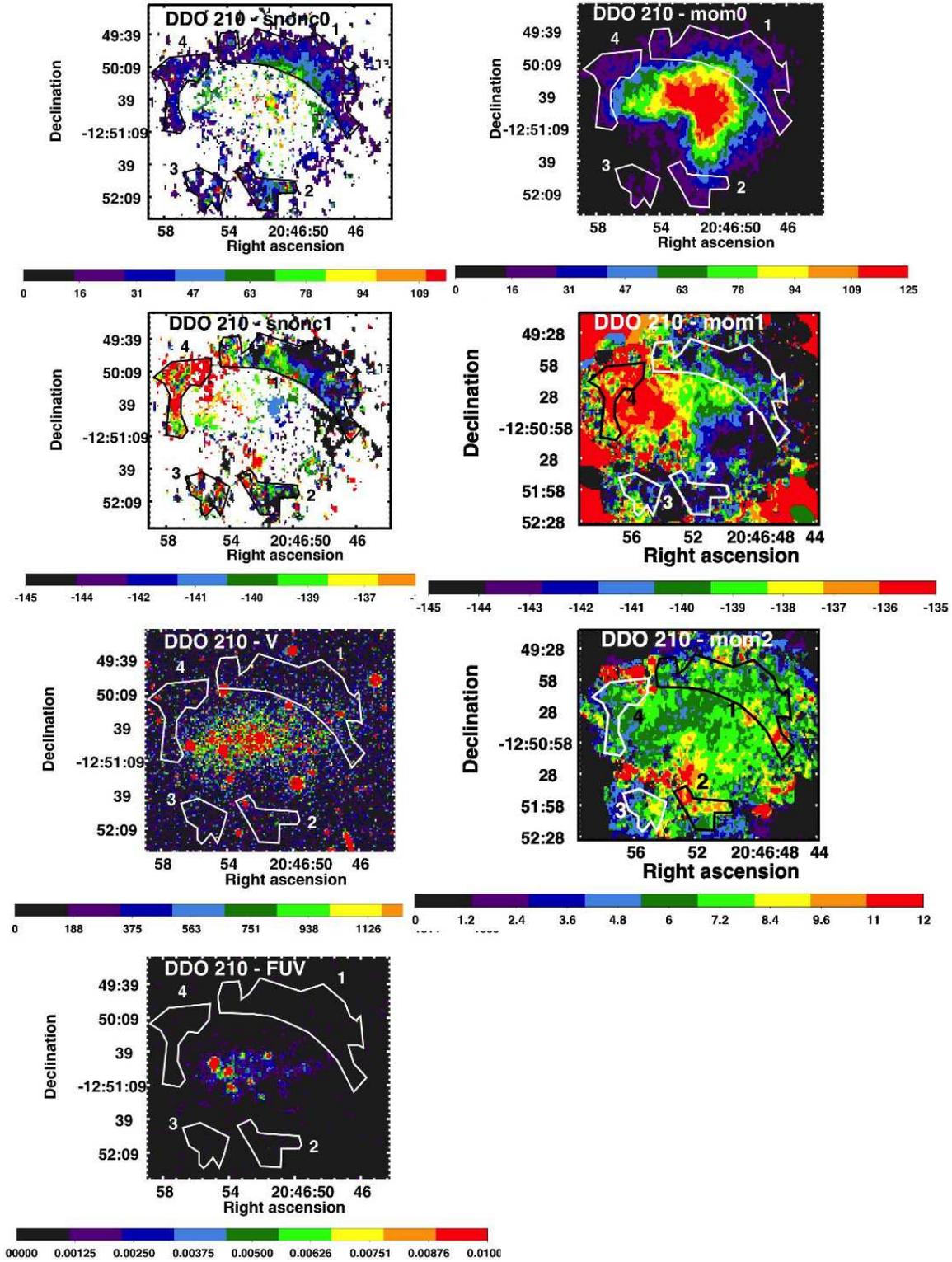}
%\vskip -0.9truein
\caption{For DDO 210, features identified on strong noncircular moment 0 (snonc0) maps
shown on snonc0, strong non-circular moment 1 (``snonc1''), total \HI\ moment 0 (``mom0''), total \HI\ moment 1 (``mom1''), total \HI\ moment 2 (``mom2''),
FUV, \ha\, and $V$-band maps.
There is no \ha\ emission in DDO 210.
Properties of the regions are given in Table \ref{tab-reg}.
Colorbar units are as in Figure \ref{fig-maps1}.
%Colorbar units are as follows: moment 0 maps are Jy beam$^{-1}$ m s$^{-1}$; snonc1 is km s$^{-1}$;
%total \HI\ moment 1 and moment 2 maps are m s$^{-1}$; FUV image is counts s$^{-1}$ 
%that can be converted to erg s$^{-1}$ cm$^{-2}$ \AA$^{-1}$ by multiplying by $1.4\times10^{-15}$;
$V$-band image is counts that can be converted to a Johnson $V$ magnitude using 
%$-2.5\log({\rm counts}/200) = V - 25 + -0.053 + 0.022\times(B-V)$ \citep{he06}.
$-2.5\log({\rm counts}/200) = V - 25.05 + 0.022\times(B-V)$ \citep{he06}.
\label{fig-maps9}}
\end{figure}

\begin{figure}
\vskip -0.95truein
\includegraphics[scale=0.825]{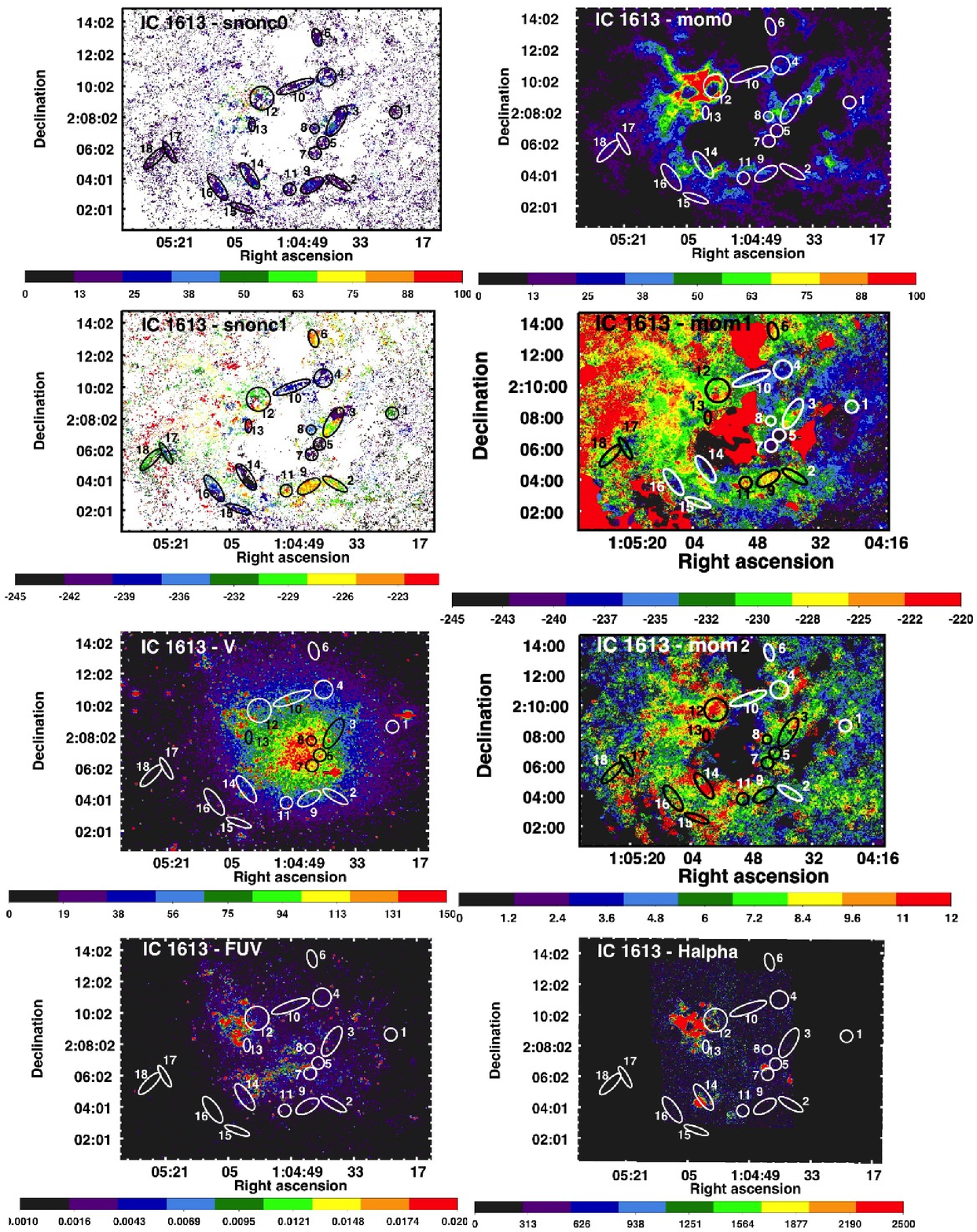}
%\vskip -0.8truein
\caption{For IC 1613, features identified on strong noncircular moment 0 (snonc0) maps
shown on snonc0, strong non-circular moment 1 (``snonc1''), total \HI\ moment 0 (``mom0''), total \HI\ moment 1 (``mom1''), total \HI\ moment 2 (``mom2''),
FUV, \ha\, and $V$-band maps.
Properties of the regions are given in Table \ref{tab-reg}.
Colorbar units are as in Figure \ref{fig-maps1}.
%Colorbar units are as follows: moment 0 maps are Jy beam$^{-1}$ m s$^{-1}$; snonc1 is km s$^{-1}$;
%total \HI\ moment 1 and moment 2 maps are m s$^{-1}$; FUV image is counts s$^{-1}$ 
%that can be converted to erg s$^{-1}$ cm$^{-2}$ \AA$^{-1}$ by multiplying by $1.4\times10^{-15}$;
\ha\ image is counts that can be converted to erg s$^{-1}$ cm$^{-2}$ by multiplying by $0.272\times10^{-18}$ \citep{he04};
and $V$-band image is counts that can be converted to a Johnson $V$ magnitude using 
%$-2.5\log({\rm counts}/600) = V - 25 + 4.023 + 0.169\times1.28 + 0.007\times(B-V)$ \citep{he06}.
$-2.5\log({\rm counts}/600) = V - 20.76 + 0.007\times(B-V)$ \citep{he06}.
\label{fig-maps10}}
\end{figure}

\begin{figure}
\vskip -0.95truein
\includegraphics[scale=0.825]{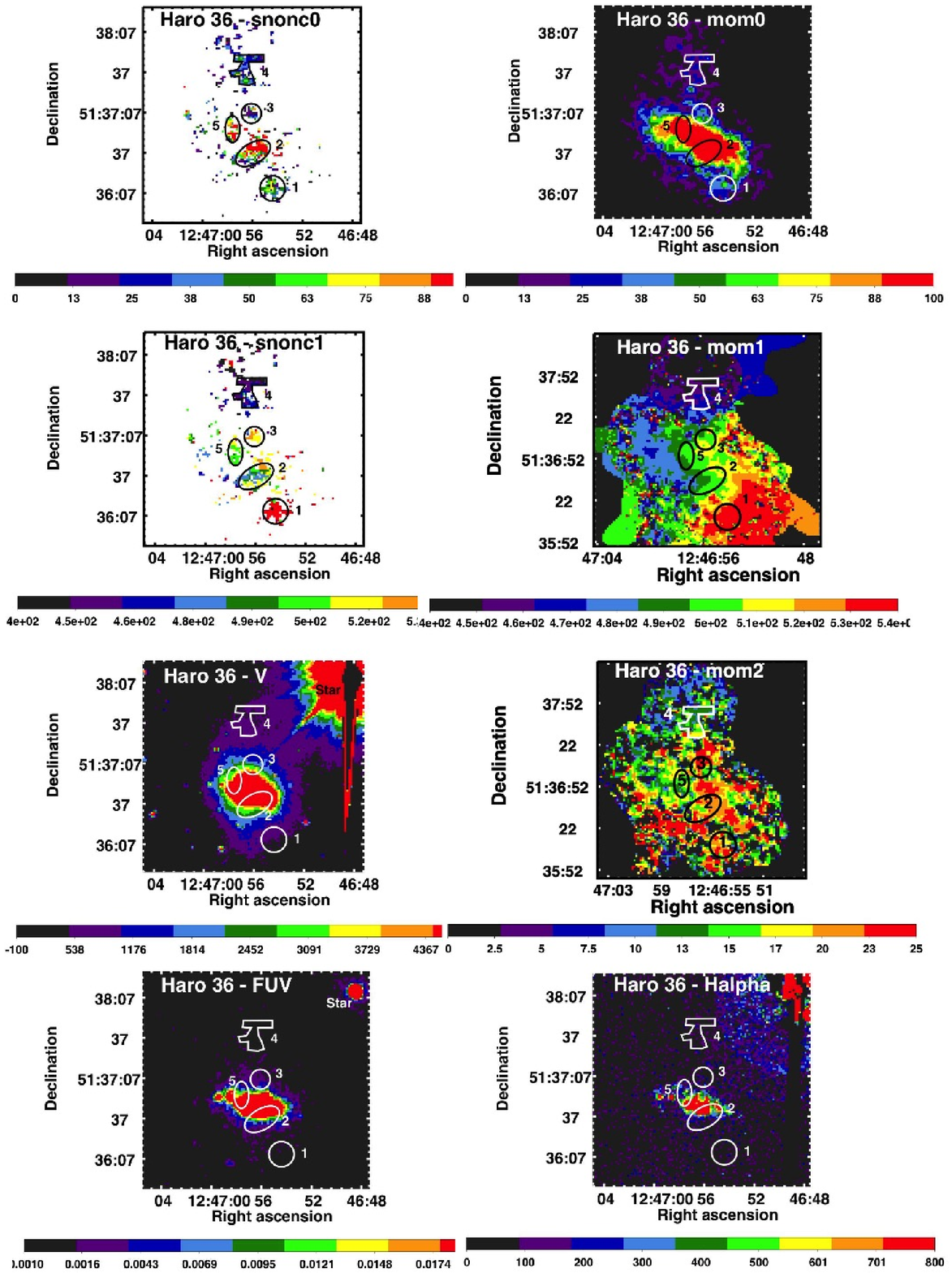}
%\vskip -0.9truein
\caption{For Haro 36, features identified on strong noncircular moment 0 (snonc0) maps
shown on snonc0, strong non-circular moment 1 (``snonc1''), total \HI\ moment 0 (``mom0''), total \HI\ moment 1 (``mom1''), total \HI\ moment 2 (``mom2''),
FUV, \ha\, and $V$-band maps.
Properties of the regions are given in Table \ref{tab-reg}.
Colorbar units are as in Figure \ref{fig-maps1}.
%Colorbar units are as follows: moment 0 maps are Jy beam$^{-1}$ m s$^{-1}$; snonc1 is km s$^{-1}$;
%total \HI\ moment 1 and moment 2 maps are m s$^{-1}$; FUV image is counts s$^{-1}$ 
%that can be converted to erg s$^{-1}$ cm$^{-2}$ \AA$^{-1}$ by multiplying by $1.4\times10^{-15}$;
\ha\ image is counts that can be converted to erg s$^{-1}$ cm$^{-2}$ by multiplying by $0.754\times10^{-18}$ \citep{he04};
and $V$-band image is counts that can be converted to a Johnson $V$ magnitude using 
%$-2.5\log({\rm counts}/900) = V - 25 + 1.532 + 0.345\times1.05 + 0.007\times(B-V)$ \citep{he06}.
$-2.5\log({\rm counts}/900) = V -23.11 + 0.007\times(B-V)$ \citep{he06}.
\label{fig-maps11}}
\end{figure}

We examined the snonc and wnonc moment 0 maps for significant features. The noncircular motion gas maps have
lots of noise, but we were looking for large, coherent regions of gas such as blobs or filaments. 
These kinematically coherent regions connect both spatially and in velocity space at least 30 pixels 
which have a signal-to-noise greater than 3. 
A spatial pixel is 1.5\arcsec\ and the channel width is either 2.58 \kms\ or 1.29 \kms, as given in Table \ref{tab-gal}.
So, for example, a region with a 6 pixel diameter would be 31 pc in diameter in IC 1613 or 450 pc in DDO 52,
and a 30 pixel area would be 1.2 times the beam area for IC 1613 and 1.7 times the beam area of DDO 52.
The regions were identified on snonc moment 0 (``snonc0'') maps 
by eye as a contiguous region of pixels.
The regions are plotted on the snonc0, snonc moment 1 (``snonc1''), 
total \HI\ moment 0 (``\HI\ mom0''),
total \HI\ moment 1 (``\HI\ mom1''), FUV, \ha, and $V$-band images of the galaxy. 
These figures are shown in Figures \ref{fig-maps1} through \ref{fig-maps11} for the 11 galaxies
in which such features were found. Galaxies without significant noncircular motion gas features have a zero in
the column for the number of regions $N_{reg}$ in Table \ref{tab-gal}.

The regions were encircled with a circle, ellipse, or polygon, and we measured the statistics of the region within that geometrical shape.
The properties of the regions are given in Table \ref{tab-reg} and include the R.A.\ and Decl.\ of the center; the major axis, minor
axis, and position angle (P.A.) of the region, if not a polygon; the encircled area in square-arcseconds; the \HI\ mass; the surface density of the gas
averaged over the region; and the difference in velocity compared to the surrounding gas in bulk ordered rotation.
The mass of the region is the mass measured in the snonc0 map and the velocity is the median in the encircled area on the snonc1 map.

% fig1

% fig2

% fig3

% fig4

% fig5

% fig6

% fig7

% fig8

% fig9

% fig10

% fig11

\clearpage

\begin{deluxetable}{lcccccccccc}
\tabletypesize{\tiny}
%\rotate
\tablecaption{Region Properties\label{tab-reg}}
\tablewidth{0pt}
\tablehead{
\colhead{Galaxy} & \colhead{Region} &
\colhead{R.A.} & \colhead{Decl.} &
\colhead{$\rm l_{maj}$\tablenotemark{a}} & \colhead{$\rm l_{min}$\tablenotemark{a}} &
\colhead{P.A.\tablenotemark{a}} & \colhead{Area} &
\colhead{log Mass} & \colhead{$\log \Sigma_{\rm HI}$\tablenotemark{b}} &
\colhead{$\Delta V$\tablenotemark{c}} \\
\colhead{} & \colhead{} &
\colhead{(hh:mm:ss.s)} & \colhead{(dd:mm:ss)} &
\colhead{(arcs)} & \colhead{(arcs)} &
\colhead{(deg)} & \colhead{(arcs$^2$)} &
\colhead{($M\solar$)} &
\colhead{($M\solar$ pc$^{-2}$)} &
\colhead{(\kms)}
}
\startdata
CVnIdwA  &  1 & 12:38:38.0 & +32:45:50 & 23.2 & 12.0 &   0 &  576 & 6.384$\pm$0.002 & 1.139$\pm$0.002 &  -2\\
         &  2 & 12:38:42.5 & +32:45:29 & 26.7 & 13.6 &  40 & 1013 & 6.571$\pm$0.002 & 1.081$\pm$0.002 &   -3\\
DDO 47   &  1 & 7:41:5 0.2 & +16:48:31 & 26.1 &  8.8 & 110 &  556 & 6.167$\pm$0.004 & 0.618$\pm$0.004 &  -6\\
         &  2 & 7:41:5 1.2 & +16:47:02 & 28.5 & 10.4 & 120 &  693 & 6.202$\pm$0.004 & 0.559$\pm$0.004 &   8\\
         &  3 & 7:41:5 3.8 & +16:47:30 & 25.9 & 13.0 &  90 &  765 & 6.357$\pm$0.003 & 0.670$\pm$0.003 &   6\\
         &  4 & 7:41:5 6.5 & +16:47:23 & 13.3 & 12.3 &  90 &  448 & 6.005$\pm$0.005 & 0.551$\pm$0.005 &  -9\\
         &  5 & 7:41:5 8.1 & +16:50:16 & 25.9 & 22.0 &  90 & 1224 & 6.571$\pm$0.003 & 0.680$\pm$0.003 &  -1\\
         &  6 & 7:41:5 8.9 & +16:48:40 & 23.3 & 20.7 &  90 & 1175 & 6.531$\pm$0.003 & 0.658$\pm$0.003 & -10\\
         &  7 & 7:42:0 1.3 & +16:49:35 & 15.6 & 14.3 &  90 &  529 & 6.167$\pm$0.004 & 0.640$\pm$0.004 &   6\\
         &  8 & 7:42:0 2.3 & +16:47:24 & 23.3 & 22.0 & 180 & 1325 & 6.805$\pm$0.002 & 0.878$\pm$0.002 &   3\\
         &  9 & 7:42:0 5.0 & +16:50:03 & 18.5 & 11.0 &  90 &  470 & 6.014$\pm$0.005 & 0.538$\pm$0.005 &   4\\
         & 10 & 7:42:0 4.9 & +16:47:25 & 17.5 &  8.1 & 180 &  347 & 6.019$\pm$0.005 & 0.676$\pm$0.005 &   5\\
         & 11 & 7:42:0 5.8 & +16:48:06 & 28.5 & 10.4 &  30 &  819 & 6.313$\pm$0.004 & 0.595$\pm$0.003 &   0\\
         & 12 & 7:42:0 6.5 & +16:49:16 & 17.5 & 13.9 & 180 &  608 & 6.062$\pm$0.005 & 0.475$\pm$0.005 &   8\\
DDO 50   &  1 & 8:19:2 0.0 & +70:44:03 & 44.5 & 13.1 &  55 & 1355 & 6.659$\pm$0.002 & 1.092$\pm$0.002 &  -9\\
         &  2 & 8:19:1 8.2 & +70:42:19 & 26.5 & 11.6 & 150 &  574 & 6.384$\pm$0.003 & 1.191$\pm$0.003 &  13\\
         &  3 & 8:19:2 3.5 & +70:42:58 & 32.5 & 13.3 &  55 &  893 & 6.592$\pm$0.002 & 1.206$\pm$0.002 &  -3\\
         &  4 & 8:19:3 5.7 & +70:43:33 & 44.5 & 13.1 & 150 & 1078 & 6.473$\pm$0.003 & 1.007$\pm$0.003 &  19\\
DDO 53   &  1 & 8:34:1 4.0 & +66:11:22 &  2.4 &  0.3 &  90 &  268 & 5.655$\pm$0.008 & 0.745$\pm$0.008 & -10\\
DDO 70   &  1 &  9:59:48.8 &  +5:19:33 & 29.2 & 15.1 & 180 &  981 & 4.903$\pm$0.003 & 0.311$\pm$0.003 &   6\\
         &  2 &  9:59:49.2 &  +5:21:26 & 17.3 & 17.3 &  90 &  824 & 5.068$\pm$0.003 & 0.552$\pm$0.003 &   6\\
         &  3 &  9:59:49.9 &  +5:23:41 & 14.0 & 11.9 &  90 &  353 & 4.382$\pm$0.006 & 0.235$\pm$0.006 &   6\\
         &  4 &  9:59:55.8 &  +5:23:24 & 14.0 & 11.9 &  90 &  394 & 4.455$\pm$0.005 & 0.259$\pm$0.005 &  -3\\
         &  5 &  9:59:57.9 &  +5:19:44 & 31.3 & 15.1 &  30 &  983 & 5.566$\pm$0.001 & 0.974$\pm$0.002 &   6\\
         &  6 &  9:59:58.1 &  +5:18:37 & 22.7 & 11.9 & 150 &  720 & 4.987$\pm$0.003 & 0.530$\pm$0.003 &   8\\
         &  7 & 10:00:01.9 &  +5:20:22 & 19.4 & 11.9 &  90 &  542 & 5.068$\pm$0.003 & 0.735$\pm$0.003 &  10\\
         &  8 & 10:00:04.6 &  +5:20:01 & 28.1 & 22.7 & 180 & 1508 & 5.492$\pm$0.002 & 0.715$\pm$0.002 &  -9\\
         &  9 & 10:00:07.5 &  +5:18:05 & 29.2 & 27.0 &  90 & 2075 & 5.378$\pm$0.002 & 0.462$\pm$0.002 &   7\\
         & 10 & 10:00:07.5 &  +5:20:29 & 14.4 &  9.6 &  30 &  297 & 4.337$\pm$0.006 & 0.264$\pm$0.006 &   6\\
         & 11 & 10:00:08.4 &  +5:20:56 & 15.1 & 11.9 & 150 &  437 & 4.525$\pm$0.005 & 0.285$\pm$0.005 & -18\\
         & 12 & 10:00:09.6 &  +5:19:34 & 23.8 & 22.7 & 180 &  972 & 5.178$\pm$0.002 & 0.591$\pm$0.002 &   2\\
         & 13 & 10:00:09.8 &  +5:23:03 & 49.6 & 21.5 & 290 & 2750 & 5.135$\pm$0.002 & 0.096$\pm$0.002 & -10\\
         & 14 & 10:00:10.3 &  +5:21:45 & 24.9 & 11.9 &  90 &  686 & 4.655$\pm$0.004 & 0.218$\pm$0.004 &   1\\
DDO 126  &  1 & 12:27:01.9 & +37:08:43 & 12.3 &  5.2 &  90 &  126 & 5.607$\pm$0.012 & 0.753$\pm$0.012 &   4\\
         &  2 & 12:27:03.9 & +37:08:28 & 14.5 &  9.6 & 140 &  371 & 6.362$\pm$0.005 & 1.040$\pm$0.005 &   5\\
         &  3 & 12:27:05.3 & +37:08:48 & 18.1 & 14.3 &  90 &  657 & 6.690$\pm$0.003 & 1.121$\pm$0.003 &   6\\
         &  4 & 12:27:08.1 & +37:07:50 & 16.8 &  8.4 &  90 &  333 & 6.276$\pm$0.005 & 1.001$\pm$0.005 &   6\\
         &  5 & 12:27:12.3 & +37:08:05 & 11.7 & 11.0 &   0 &  275 & 5.814$\pm$0.009 & 0.650$\pm$0.009 &  -6\\
         &  6 & 12:27:12.4 & +37:07:09 & 10.4 & 10.4 &  90 &  257 & 5.922$\pm$0.008 & 0.761$\pm$0.008 &  -2\\
DDO 133  &  1 & 12:32:50.9 & +31:34:52 & 19.4 & 11.9 &  30 &  576 & 5.232$\pm$0.007 & 0.014$\pm$0.007 &  -5\\
         &  2 & 12:32:48.4 & +31:32:49 & 12.9 & 12.9 &   0 &  329 & 5.510$\pm$0.005 & 0.535$\pm$0.005 &   1\\
         &  3 & 12:32:50.7 & +31:32:41 & 15.8 & 15.8 &   0 &  540 & 5.848$\pm$0.003 & 0.655$\pm$0.003 &  26\\
         &  4 & 12:32:51.3 & +31:32:13 & 12.9 & 12.9 &   0 &  410 & 5.916$\pm$0.003 & 0.844$\pm$0.003 &   4\\
         &  5 & 12:32:52.4 & +31:31:47 & 13.2 &  9.9 &  40 &  230 & 5.372$\pm$0.006 & 0.552$\pm$0.006 & -20\\
         &  6 & 12:32:53.2 & +31:32:21 & 16.2 &  7.6 &  90 &  214 & 5.731$\pm$0.004 & 0.941$\pm$0.004 &   2\\
         &  7 & 12:32:54.1 & +31:32:43 & 28.1 &  8.6 &  90 &  558 & 6.081$\pm$0.003 & 0.874$\pm$0.003 &  -8\\
         &  8 & 12:32:54.6 & +31:31:33 & 24.8 &  9.7 &   0 &  430 & 5.887$\pm$0.017 & 0.794$\pm$0.003 &   2\\
         &  9 & 12:32:55.6 & +31:33:50 & 25.9 & 13.0 & 110 &  837 & 5.744$\pm$0.004 & 0.363$\pm$0.004 &   -4\\
         & 10 & 12:32:55.5 & +31:30:41 & 12.9 & 12.9 &   0 &  304 & 5.378$\pm$0.006 & 0.435$\pm$0.006 &  -3\\
         & 11 & 12:32:56.4 & +31:29:48 & 19.4 &  7.6 &   0 &  360 & 5.317$\pm$0.006 & 0.302$\pm$0.006 &   4\\
         & 12 & 12:32:58.7 & +31:31:41 & 16.9 & 16.9 &   0 &  527 & 5.798$\pm$0.004 & 0.618$\pm$0.004 &   2\\
         & 13 & 12:32:59.2 & +31:32:42 & 28.9 & 28.9 &   0 & 1465 & 6.104$\pm$0.003 & 0.478$\pm$0.003 &   1\\
         & 14 & 12:32:59.8 & +31:34:19 & 14.0 &  8.6 & 110 &  248 & 5.500$\pm$0.005 & 0.646$\pm$0.005 &  -1\\
         & 15 & 12:32:59.4 & +31:33:39 & 15.1 &  8.6 & 140 &  284 & 5.439$\pm$0.005 & 0.525$\pm$0.005 &  -5\\
         & 16 & 12:33:01.2 & +31:33:22 & 15.2 & 15.2 &   0 &  410 & 5.694$\pm$0.004 & 0.622$\pm$0.004 &   2\\
         & 17 & 12:33:04.5 & +31:32:05 & 27.0 & 22.7 &  35 & 1719 & 6.123$\pm$0.002 & 0.428$\pm$0.002 &   13\\
         & 18 & 12:33:02.3 & +31:30:32 & 59.4 & 18.1 &  30 & 2698 & 6.134$\pm$0.002 & 0.243$\pm$0.003 &   10\\
DDO 168  &  1 & 13:14:24.6 & +45:54:30 & 13.0 & 11.9 & 180 &  317 & 6.094$\pm$0.005 & 0.955$\pm$0.005 &  -5\\
         &  2 & 13:14:25.5 & +45:55:34 & 14.0 & 10.8 & 180 &  340 & 6.351$\pm$0.004 & 1.182$\pm$0.004 &  -8\\
         &  3 & 13:14:27.4 & +45:55:30 &  9.2 &  9.2 &   0 &  144 & 6.392$\pm$0.004 & 1.595$\pm$0.004 &   8\\
         &  4 & 13:14:27.8 & +45:54:19 & 18.4 & 15.1 & 180 &  585 & 6.576$\pm$0.003 & 1.170$\pm$0.003 &   -8\\
         &  5 & 13:14:28.0 & +45:53:50 & 14.0 &  9.7 &  90 &  254 & 5.815$\pm$0.007 & 0.773$\pm$0.007 &   -11\\
DDO 210  &  1 & 20:46:49.1 & -12:50:11 &  0.0 &  0.0 &   0 & 3992 & 5.371$\pm$0.002 & 0.489$\pm$0.002 &  -3\\
         &  2 & 20:46:51.4 & -12:52:00 &  0.0 &  0.0 &   0 &  963 & 4.761$\pm$0.004 & 0.497$\pm$0.004 &  -6\\
         &  3 & 20:46:55.1 & -12:52:08 &  0.0 &  0.0 &   0 &  576 & 4.540$\pm$0.005 & 0.500$\pm$0.005 &   2\\
         &  4 & 20:46:56.7 & -12:50:31 &  0.0 &  0.0 &   0 & 1346 & 4.793$\pm$0.003 & 0.385$\pm$0.004 &   2\\
Haro 36  &  1 & 12:46:54.4 & +51:36:10 &  9.4 &  9.4 &   0 &  155 & 6.478$\pm$0.008 & 0.979$\pm$0.008 &   6\\
         &  2 & 12:46:55.9 & +51:36:36 & 14.3 &  7.7 & 120 &  236 & 6.955$\pm$0.005 & 1.274$\pm$0.005 &  -5\\
         &  3 & 12:46:56.0 & +51:37:06 &  7.3 &  7.3 &   0 &   74 & 5.952$\pm$0.015 & 0.772$\pm$0.015 &   8\\
         &  4 & 12:46:56.6 & +51:37:39 &  0.0 &  0.0 &   0 &  230 & 6.546$\pm$0.007 & 0.875$\pm$0.007 &  14\\
         &  5 & 12:46:57.5 & +51:36:54 &  9.9 &  5.4 &   0 &   95 & 6.528$\pm$0.008 & 1.244$\pm$0.007 &  20\\
IC 1613  &  1 &  1:04:23.4 &  +2:08:38 & 23.8 & 23.8 &   0 &  833 & 4.369$\pm$0.006 & 0.387$\pm$0.006 &   7\\
         &  2 &  1:04:37.9 &  +2:04:12 & 54.0 & 16.2 &  60 & 1748 & 4.728$\pm$0.004 & 0.422$\pm$0.004 &   9\\
         &  3 &  1:04:38.4 &  +2:08:12 & 63.7 & 26.4 & 150 & 3929 & 5.308$\pm$0.003 & 0.651$\pm$0.002 &   7\\
         &  4 &  1:04:40.9 &  +2:11:01 & 34.9 & 34.9 &   0 & 1829 & 4.930$\pm$0.003 & 0.606$\pm$0.003 &  -3\\
         &  5 &  1:04:41.9 &  +2:06:50 & 23.7 & 23.7 &   0 &  936 & 4.517$\pm$0.005 & 0.483$\pm$0.005 &  -7\\
         &  6 &  1:04:43.4 &  +2:13:28 & 34.2 & 18.0 &  15 & 1361 & 4.646$\pm$0.004 & 0.453$\pm$0.004 &   4\\
         &  7 &  1:04:43.9 &  +2:06:10 & 23.7 & 23.7 &   0 &  837 & 4.484$\pm$0.005 & 0.500$\pm$0.005 &  -3\\
         &  8 &  1:04:44.0 &  +2:07:45 & 18.2 & 18.2 &   0 &  578 & 4.447$\pm$0.006 & 0.623$\pm$0.005 &  -4\\
         &  9 &  1:04:44.6 &  +2:04:04 & 48.6 & 23.4 & 120 & 3193 & 5.143$\pm$0.003 & 0.575$\pm$0.002 &  10\\
         & 10 &  1:04:48.9 &  +2:10:26 & 75.7 & 28.5 & 110 & 1960 & 4.851$\pm$0.003 & 0.500$\pm$0.003 &  -9\\
         & 11 &  1:04:50.4 &  +2:03:48 & 23.7 & 23.7 &   0 & 1024 & 4.567$\pm$0.005 & 0.494$\pm$0.005 &   9\\
         & 12 &  1:04:57.5 &  +2:09:41 & 45.4 & 45.4 &   0 & 2970 & 5.359$\pm$0.002 & 0.825$\pm$0.002 &  -2\\
         & 13 &  1:05:00.0 &  +2:07:58 & 25.2 & 12.6 &   0 &  626 & 3.357$\pm$0.006 & 0.500$\pm$0.006 &   9\\
         & 14 &  1:05:00.6 &  +2:04:40 & 59.4 & 22.1 &  35 & 2732 & 5.122$\pm$0.003 & 0.623$\pm$0.002 & -14\\
         & 15 &  1:05:02.5 &  +2:02:32 & 50.4 & 12.6 &  70 & 1244 & 4.653$\pm$0.004 & 0.494$\pm$0.004 &  -8\\
         & 16 &  1:05:08.8 &  +2:03:51 & 59.4 & 22.1 &  35 & 2716 & 5.029$\pm$0.003 & 0.531$\pm$0.003 &  -9\\
         & 17 &  1:05:21.0 &  +2:05:59 & 46.7 & 12.5 &  30 & 1298 & 4.631$\pm$0.004 & 0.459$\pm$0.004 &  -6\\
         & 18 &  1:05:24.8 &  +2:05:31 & 56.0 & 17.8 & 135 & 1910 & 4.758$\pm$0.004 & 0.415$\pm$0.004 &  -5\\
\enddata
\tablenotetext{a}{~Length of major axis, length of minor axis, and position angle of region for those outlined with an ellipse.
For those outlined with a circle, the major axis and minor axis are the same and the P.A.\ is given as 0 degrees.
For those outlined with a polygon, zeroes are entered for all three values.}
\tablenotetext{b}{Surface mass density determined from the mass divided by the area.}
\tablenotetext{c}{Median velocity in the encircled region minus the velocity of the gas engaged in bulk motions
as sampled adjacent to the snonc0 region of interest.
Note that our ability to measure the velocity of features and the velocity offset is limited by the fact that the features 
are extended and cover a range of velocities.
The velocities of the regions in DDO 210 are particularly uncertain because of the limited rotation and the location of the
features in the outer edges of the galaxy.} 
\end{deluxetable}

\clearpage
\section{Results} \label{sec-results}

\subsection{Streaming motions} \label{sec-streaming}

\subsubsection{DDO 133}

% fig12
\begin{figure}[t!]
%\vskip -1.0truein
%\includegraphics[scale=0.4]{/ta/reu18/analysis/fig/d133/falsecolorubv3_cut.eps}
\includegraphics[scale=0.8]{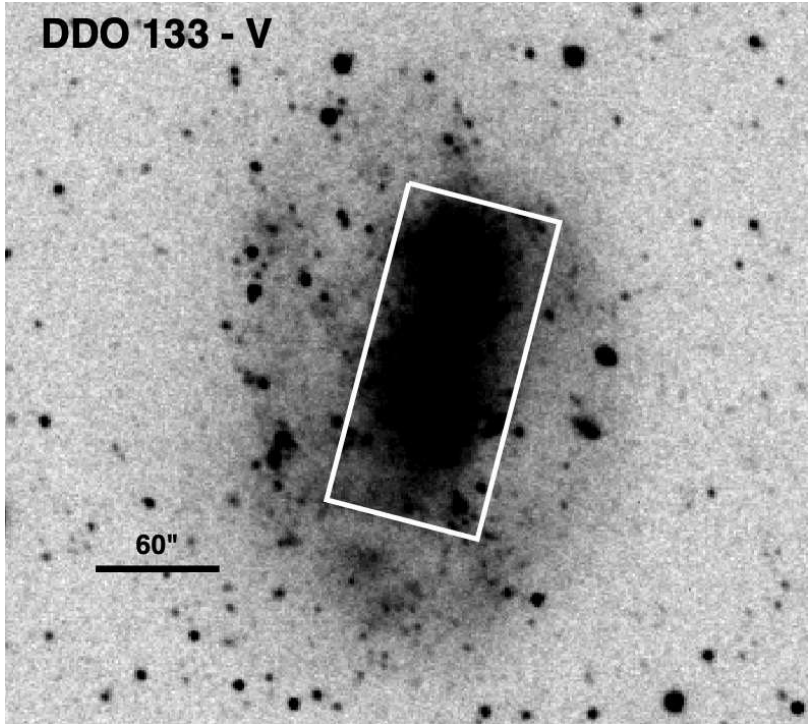}
%\vskip -1.0truein
\caption{$V$-band image of DDO 133 to show the stellar bar structure which is outlined with the white box.
\label{fig-d133ubv}}
\end{figure}

\begin{figure}[t!]
\vskip -0.85truein
\begin{center}
\includegraphics[scale=0.95]{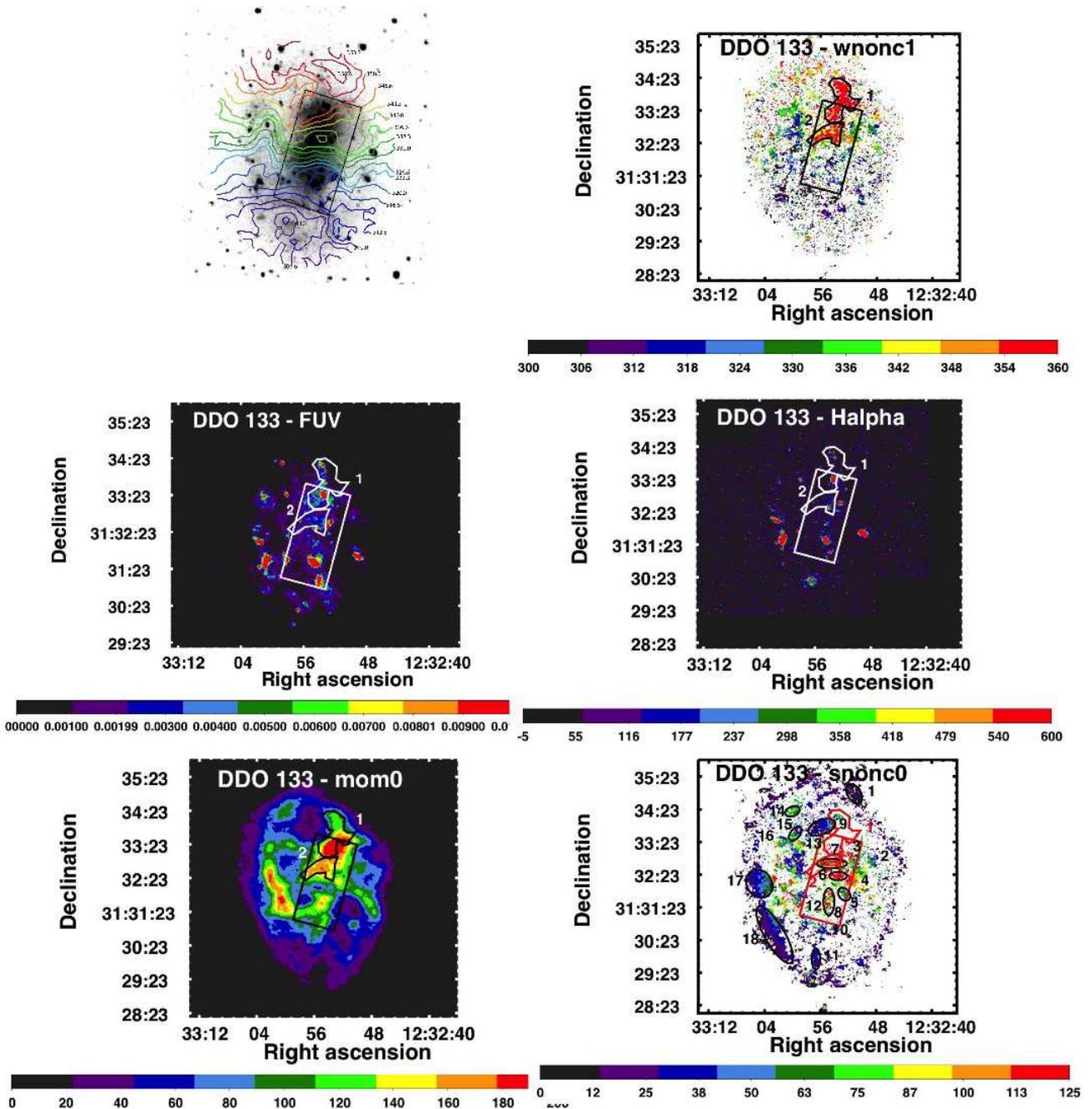}
\end{center}
%\vskip -3.15truein
\caption{DDO 133 weak noncircular (wnonc) regions marked on various images of the galaxy.
Upper left: $V$-band image with contours of the \HI\ velocity field. 
The black rectangle outlines the edges of the stellar bar as seen in a deep display of the $V$ image.
The velocity contours are labeled in \kms. 
The ``crinkle'' pattern in the velocity contours due to streaming motions around the bar is evident.
Other panels: The underlying image is given in the title. The bar and two wnonc regions are outlined.
Bottom right: The snonc and wnonc regions and bar are outlined on the snonc moment 0 map to show
the relationship between of some of the snonc regions with the bar and wnonc regions.
The units of the colorbars are the same as those in Figure \ref{fig-maps7}.
\label{fig-d133bar}}
\end{figure}

In optical images, DDO 133 appears to have a stellar bar. This is visible as a rotation in the position angle of isophotes between the
inner galaxy and the outer. In the $V$-band image shown in Figure \ref{fig-d133ubv} one can see the bright, boxy inner structure 
that is the stellar bar and the more diffuse oval structure that is the stellar disk.
The bar is outlined on other images in Figure \ref{fig-d133bar}.
Here we have included a velocity map of the weak noncircular gas map (wnonc1) where we have identified two
wnonc regions that are coincident with the northern end of the bar (wnonc region 1) and upper part of the body of the bar (wnonc region 2).
Properties of the two wnonc regions and the bar are given in Table \ref{tab-d133bar}.
The wnonc regions in DDO 133 are located on a ridge of \HI\ in the integrated moment 0 map.
Furthermore, there is a clump that is bright in \ha\ and FUV associated with wnonc region 1 at the end of the stellar bar.

\clearpage

% fig13

\clearpage

\begin{deluxetable}{lccccccc}
\tabletypesize{\tiny}
%\rotate
\tablecaption{DDO 133 Weak noncircular Regions and Bar\tablenotemark{a} \label{tab-d133bar}}
\tablewidth{0pt}
\tablehead{
\colhead{Region} &
\colhead{R.A.} & \colhead{Decl.} &
\colhead{$\rm l_{maj}$} & \colhead{$\rm l_{min}$} &
\colhead{P.A.} & \colhead{Area} &
\colhead{$\Delta V$} \\
\colhead{} &
\colhead{(hh:mm:ss.s)} & \colhead{(dd:mm:ss)} &
\colhead{(arcs)} & \colhead{(arcs)} &
\colhead{(deg)} & \colhead{(arcs$^2$)} &
\colhead{(\kms)}
}
\startdata
wnonc 1\tablenotemark{b} & 12:32:52.8 & +31:33:38 &  \nodata    &   \nodata  &  \nodata   &   1818 &  7 \\
wnonc 2\tablenotemark{b} & 12:32:54.8 & +31:32:36 &  \nodata   &  \nodata  &  \nodata    &   1177 & 17 \\
Bar          & 12:32:54.3 & +31:32:15 & 158.5 & 75.5 & 165 & 11972 &  \nodata  \\
\enddata
\tablenotetext{a}{Parameters are defined in Table \ref{tab-reg}.}
\tablenotetext{b}{The wnonc regions are outlined with polygons and so $\rm l_{maj}$, $\rm l_{min}$,
and P.A. have no values.}
\end{deluxetable}

In the upper left panel of Figure \ref{fig-d133bar}, velocity contours (in rainbow colors) have been placed on the $V$-band image of DDO 133.
Note that along the outside edges of the bar (outlined in black), there is a
crinkle pattern in the velocity contours.
The same crinkle pattern along the edges of the bar are also seen by \citet{lin13} in the velocity field of the spiral galaxy NGC 1097.
Their hydrodynamical simulation reproduces the velocity field pattern
through streaming motions of the gas along the bar.
Furthermore, it seems likely that the star formation at the end of the stellar bar
is a result of gas piling up at the ends from streaming motions around the bar, and that this is 
also the cause of the wnonc motions there. 
The buildup of gas at the end of the bar could also have been influenced by the formation of an \HI\ hole, 
which might once have been an expanding bubble, to the east of the northern end of the bar.
This hole is evident as the circular purple region at the northern end of the mom0 image in the bottom left panel of Figure \ref{fig-d133bar}.

\subsubsection{DDO 47}

\begin{figure}[t!]
\vskip -0.95truein
\includegraphics[scale=0.7]{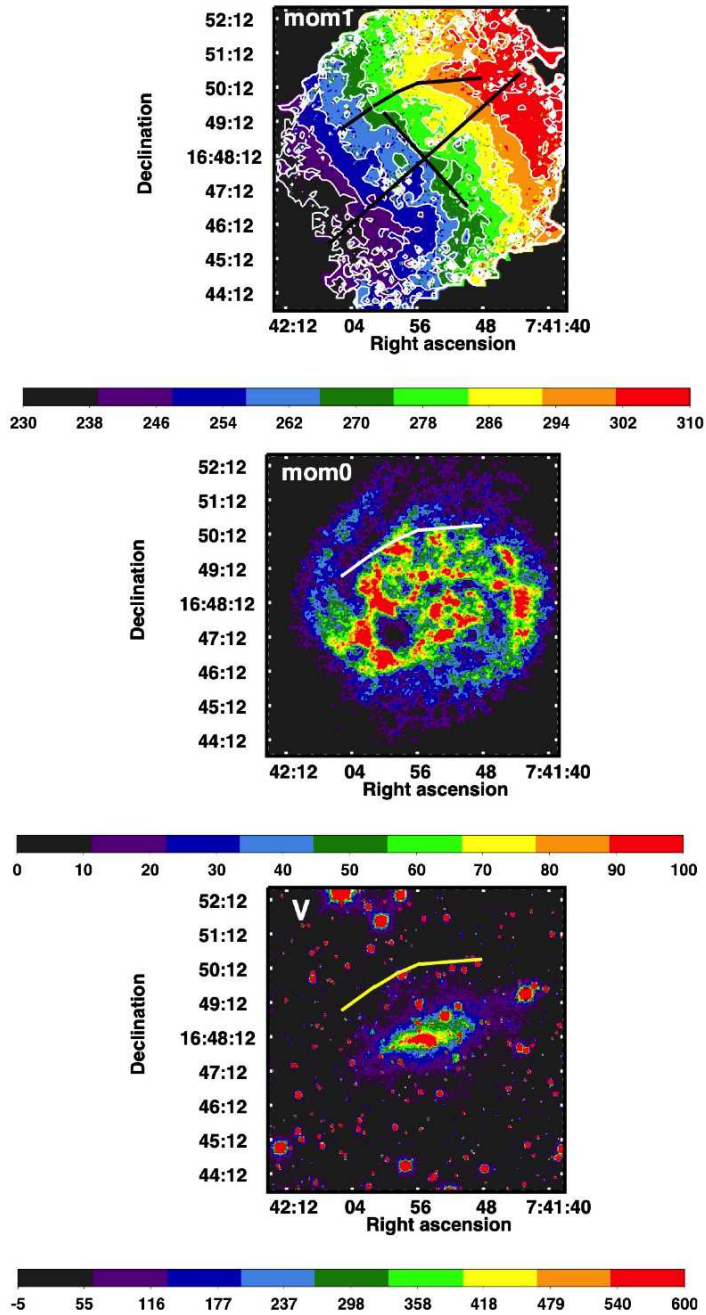}
\vskip -0.2truein
\caption{DDO 47. The solid arc in each panel traces the crinkle pattern in the velocities.
Top: False-color image velocity field (mom1). The long black solid straight line and shorter line perpendicular to that
mark the major and minor axes of the galaxy from fits to the \HI\ velocity field by \citet{oh15}. 
Ten white velocity contours from 230 \kms\ to 310 \kms\ are superposed.
Color bar units are \kms.
Middle: False-color image of the integrated \HI\ intensity (mom0).
Colorbar units are Jy beam$^{-1}$ m s$^{-1}$.
Bottom: False-color $V$-band image 
Colorbar units are the same as in Figure \ref{fig-maps2}.
A stellar bar is not obvious; the small structure in the center (colored red) is largely associated with star formation 
(see FUV image in Figure \ref{fig-maps2}).
\label{fig-d47crinkle}}
\end{figure}

DDO 47 also seems to have a crinkle pattern in the \HI\ velocity contours as shown in 
the velocity field (mom1) image in Figure \ref{fig-d47crinkle}.
However, unlike DDO 133 a stellar bar is not obvious. 
There is a small elongated structure (red in $V$ image in Figure \ref{fig-d47crinkle}) near the center of the galaxy,
but it is associated with star formation in the FUV image in Figure \ref{fig-maps2} and it is not clear that it is a bar structure.
According to \citet{bosma81}, kinematic patterns like that due to a bar can also be the result of warps in the \HI\ disk.

% fig14

Alternatively, in a study of the gas structure in DDO 47, \citet{gentile05} suggest that the
filamentary structure seen in the total \HI\ is a spiral structure.
In fact, the integrated \HI\ moment 0 map (red in top panel of Figure \ref{fig-d47crinkle})
has the general appearance of a flocculent spiral structure.
One can see that the crinkles seem
to follow the density patterns of the ``arms'' (see Section \ref{sec-filaments}).
While dwarf galaxies do not rotate fast
enough to sustain proper spiral arms \citep{gallagher84}, it is possible for
arm-like modes of instabilities to form. 
However, in dwarfs these instabilities would be transient and likely
of a shorter timescale than the rotational period of the galaxy; thus they would break
up swiftly and appear more messy than traditional arms \citep{levine94}.
We could be seeing DDO 47 in the midst of one of these transient states. 

\begin{figure}[t!]
%\vskip -1.0truein
%\includegraphics[scale=0.7]{/ta/reu18/analysis/darkmatter_3_cut.eps}
\includegraphics[scale=0.7]{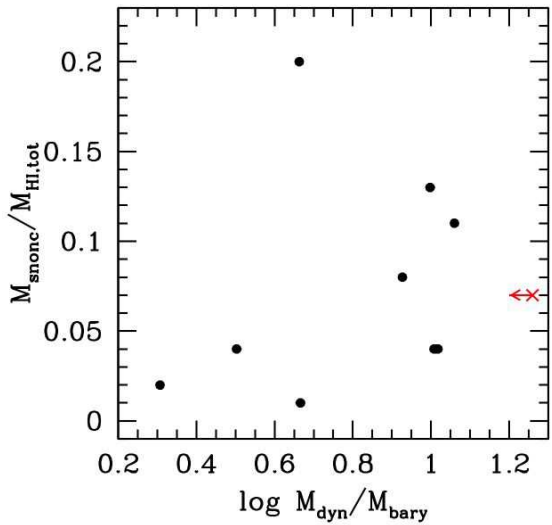}
\vskip -0.25truein
\caption{Fraction of \HI\ gas engaged in strong non-circular motions vs.\  the log of the ratio of dynamical mass to baryonic mass
$M_{dyn}/M_{bary}$ for each galaxy with analyzed snonc motion regions.
The dynamical mass comes from rotation curve analysis by \citet{oh15} and includes all forms of mass.
DDO 47 is shown as the red X. The upper limit on $M_{dyn}/M_{bary}$ for DDO 47 is due to the fact that we do not
have a stellar mass for DDO 47.
\label{fig-darkmatter}}
\end{figure}

Another consideration is that these instability modes can be dramatically influenced by the dark matter
distribution in the galaxy. The total mass to stellar mass ratio in the central region of a
galaxy compared to that in the disk influences what instability modes can be supported or if
the modes will be damped and dispersed \citep{boldrini18}.
We do not have the total mass to stellar mass ratio in the central region, but
we do have the ratio of the dynamical mass to the baryonic mass from \citet{oh15}.
We use \HI\ plus stars instead of stellar mass alone because the baryonic mass of dIrrs is dominated by the gas.
We plot the fraction of gas engaged in snonc features against the ratio of the dynamical mass to the baryonic mass $M_{dyn}/M_{bary}$
in Figure \ref{fig-darkmatter}. DDO 47 is plotted as a red X and is shown as an upper limit in
$M_{dyn}/M_{bary}$ because we do not have the stellar mass for this galaxy.
For a reasonable fraction of stellar mass, of the galaxies we can plot here, DDO 47 does have a high 
$M_{dyn}/M_{bary}$.
Furthermore, there is a suggestion of a trend of higher fraction of gas engaged in non-circular motions for higher 
$M_{dyn}/M_{bary}$, although with a lot of scatter.

We looked for the characteristic crinkle pattern in the moment 1 maps of other LITTLE THINGS galaxies that 
had been identified as possibly barred from twisting of isophotes in $V$-band images
(DDO 43, DDO 70, DDO 154, F564-V3, NGC 3738, WLM, Haro 36) as well as the rest of the sample in which bars were not suspected since weak bars are found in a large fraction of cosmological simulations of dwarfs \citep{marasco18}.
We did not find the crinkle pattern in the velocity field that would be evidence of gas streaming motions in any of these galaxies.

% fig15

\subsection{Large-scale filaments and accretion} \label{sec-filaments}

Simulations suggest that cold accretion of gas from the cosmic web onto galaxies is an on-going process
even today and a driver for on-going star formation in spirals
\citep[e.g.,][]{finlator08,forbes14,sanchez14}.
For example, \citet{schmidt16} found radial mass fluxes in 5 spiral galaxies from The \HI\ Nearby Galaxy Survey \citep[THINGS;][]{things}
that were comparable to their star formation rates.
\citet{sanchez15} find that large regions of star formation in tadpole galaxies are
significantly lower in metallicity than the rest of the galaxy. They argue that these are possibly regions 
where accreting gas has been compressed as it enters the disk and a large star-forming event was triggered.
Since the low metallicity gas rotates through the
galaxy and dissipates in less than an orbital period, the accretion events must have
happened recently for unusually low metallicity regions to be found.

Motivated by these observations, \citet{ceverino16} have run 
cosmological zoom-in simulations of the growth of a dIrr galaxy, tracking gas accretion and the
distribution of star formation and metallicity
in order to make connections between gas inflows and low metallicity regions in star forming galaxies. 
They found that a number of their simulated galaxies have clumpy \ha\ regions
with lower metallicity than the rest of the galaxy, such as observed in the tadpole galaxies. 
In the simulations, the velocities of the incoming streams of gas differ significantly from the
bulk motion of the galaxy, and terminate at the clumpy \ha\ regions.
This implies that the incoming gas streams are piling up and driving the high star
formation rate, low metallicity pockets.
To look for accretion events, therefore, we might expect to find incoming streams of gas
that are more dense and moving at a different velocity than the bulk rotation of the existing \HI\ disk.
Furthermore, impact of gas clouds on dIrrs has been observed to result in
the formation of super star clusters \citep[for example in NGC 1569 and NGC 5253,][]{johnson13,turner15}.
Thus, we might expect to find significant star-forming events to be associated with these streams.

\begin{figure}[t!]
\vskip -0.5truein
\includegraphics[scale=0.6]{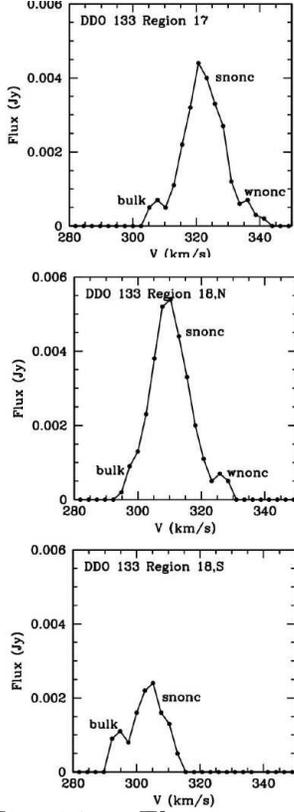}
\vskip -0.3truein
\caption{Flux versus velocity through the \HI\ cube of DDO 133 in a box 15\arcsec$\times$12\arcsec\ at the center of region 17,
in the middle of the northern half of region 18, and in the middle of the southern half of region 18.
These regions are identified on various images in Figure \ref{fig-maps7}.
The highest peak in each panel is the snonc velocity component, and the 
peak due to the velocity of gas in bulk motion in that location is labeled.
\label{fig-d133fil}}
\end{figure}

Large filamentary structures were identified in two of our galaxies: DDO 47 and DDO 133.
The filament in DDO 133 is region 18 in Figure \ref{fig-maps7}, and region 17 just above region 18, 
although somewhat detached, could also be part of this filament.
There are $2.7\times10^6$ M\solar\ of gas in this filament, which is 3\% of the total \HI\ mass of the galaxy.
The width of the filament is about 590 pc, 
the average column density of the \HI\ is $1.8\times10^{20}$ \coldens,
and the distance of the filament from the center of the galaxy in the face-on plane of the galaxy
is about 3 kpc.
As an example, Figure \ref{fig-d133fil} shows cuts through the \HI\ cube at three positions in the filament:
the middle of region 17, the middle of the north half of region 18, and the middle of the south half of region 18.
The intensity peaks due to the bulk and snonc motion gas are marked, and there is also a wnonc component clearly visible in two of the cuts.
The offset in velocity between the filament and the bulk motion gas is of order 10 \kms.
There is no connection between young regions seen in FUV or \ha\ and the filament.

% fig16

\begin{figure}[t!]
\vskip -0.5truein
\includegraphics[scale=0.4]{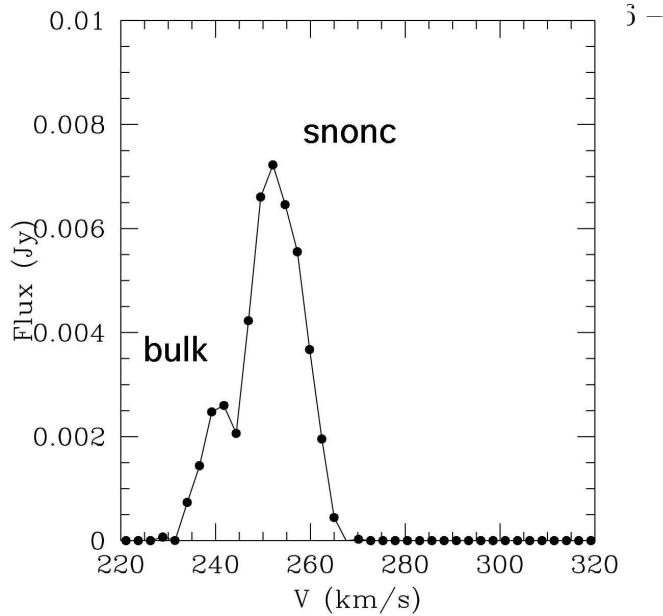}
\vskip -0.25truein
\caption{Flux versus velocity through the \HI\ cube of DDO 47 in a small box at the center of region 11.
This region is identified on various images in Figure \ref{fig-maps2}.
The taller peak is the snonc velocity,
while the shorter peak is the velocity of gas in bulk motion in that location.
\label{fig-d47fil}}
\end{figure}

The filament in DDO 47 is less coherent and is delineated by regions 9-12 (and possibly region 8) in Figure \ref{fig-maps2}.
The gas in the filament is $5.3\times10^6$ M\solar, which is 1\% of the total \HI\ mass of the galaxy, and
has median velocities 4-8 \kms\ higher than in the nearby bulk rotation gas.
An example of a cut through the \HI\ cube at a point in the middle of region 11 is shown in Figure \ref{fig-d47fil}.
The gas in bulk motion
at that location is moving around 240 \kms\ along the line of sight, while there is even
more gas there moving in noncircular motion at about 253 \kms.
The width of the filament is about 760 pc, 
the average column density of the \HI\ is $4.7\times10^{20}$ \coldens,
and the distance of the filament from the center of the galaxy in the plane of the galaxy
is about 4 kpc.
As for DDO 133, there is no connection between young regions seen in FUV or \ha\ and the filament.

% fig17

We examined the timescales for destruction of the filaments in DDO 133 and DDO 47.
We first look at the rate of growth of shearing perturbations, given by \citet{hunter98} as
$\pi G \Sigma_g/c$, where $G$ is the gravitational constant, $\Sigma_g$ is the gas surface density and $c$ is the gas velocity dispersion.
We multiply the \HI\ surface density by 1.34 to include Helium. 
For the filament in DDO 133 we find that the growth rate is $5\times10^{-18}$ s$^{-1}$, which gives
a timescale of $6\times10^9$ yr. Thus, it would take a very long time to destroy the filament by the 
growth of a shearing perturbation, if the filament is in the plane of the galaxy disk.

Next we consider the time for the filament to disperse, given the width and the velocity dispersion of the gas.
The timescale to disperse (width divided by velocity dispersion) is $8\times10^7$ yr, which is 19\% of an orbital time for the galaxy.
This implies that dissipation may be the primary destruction mechanism for the filament in DDO 133.
This is the predominant mechanism for destruction found in the simulations of \citet{ceverino16}.

For the filament in DDO 47 we find the timescale for growth of shearing perturbations is $8\times10^7$ yr
and the timescale for dispersion is similar at $1\times10^8$ yr. The latter is 31\% of an orbital time.
Thus the filament in DDO 47 could be destroyed by either shearing perturbations or dissipation.

Finally, we consider whether the filament in DDO 133 is unstable to gravitational instabilities that might enable it
to collapse into star forming clouds. The \HI\ plus He gas density is $2.4\times 10^{20}$ atoms cm$^{-2}$.
This gas density is 20\% of the \citet{toomre64} critical gas density for instability at that radius in DDO 133 \citep{eh15}.
A typical region in the filament in DDO 47 has an \HI$+$He column density of $6.3\times10^{20}$ atoms cm$^{-2}$,
and this is 47\% of the Toomre critical gas density.
Therefore, the filaments are stable against internal collapse into self-gravitating clouds.

As to whether either of these filaments could be cosmic cold accretion, we cannot tell for sure.
However, the filaments that we observe are fairly well behaved, being nearly engaged in the
rotation pattern of the galaxy.
%The filament in DDO 47 is 3-8 \kms\
%different from the surrounding bulk motions and that in DDO 133 is indistinguishable.
By contrast, a cloud of gas falling onto the BCD VIIZw 403 has a very complex kinematic
pattern relative to the body of the galaxy \citep[see Figure 17 in][]{ashley17}.
The simulations of cold accretion in dwarf galaxies by \citet{ceverino16} (see their Figure 4) also show significant
deviations in the velocities of the infalling gas relative to the galaxy. 
Furthermore, there does not seem to be consequences of these filaments to the galaxies in terms of star formation.
While the observational signatures of accretion might depend on the details of the infall
and on the density of the gas it is falling onto,
it seems that these filaments, whatever their nature, are not having much of an impact
on their host galaxies. By the same argument and the fact that these galaxies were chosen to be fairly isolated, 
it seems unlikely that these filaments are the consequence of a recent interaction with another galaxy.
However, transient instability modes, as discussed for DDO 47 above, are possibilities.

\subsection{Relationship of noncircular motion gas to star formation}

\begin{figure}[t!]
%\vskip -0.5truein
%\includegraphics[scale=0.825]{/ta/reu18/analysis/sfr_fract_cut.eps}
%\includegraphics[scale=0.825]{fig18_1.eps}
\includegraphics{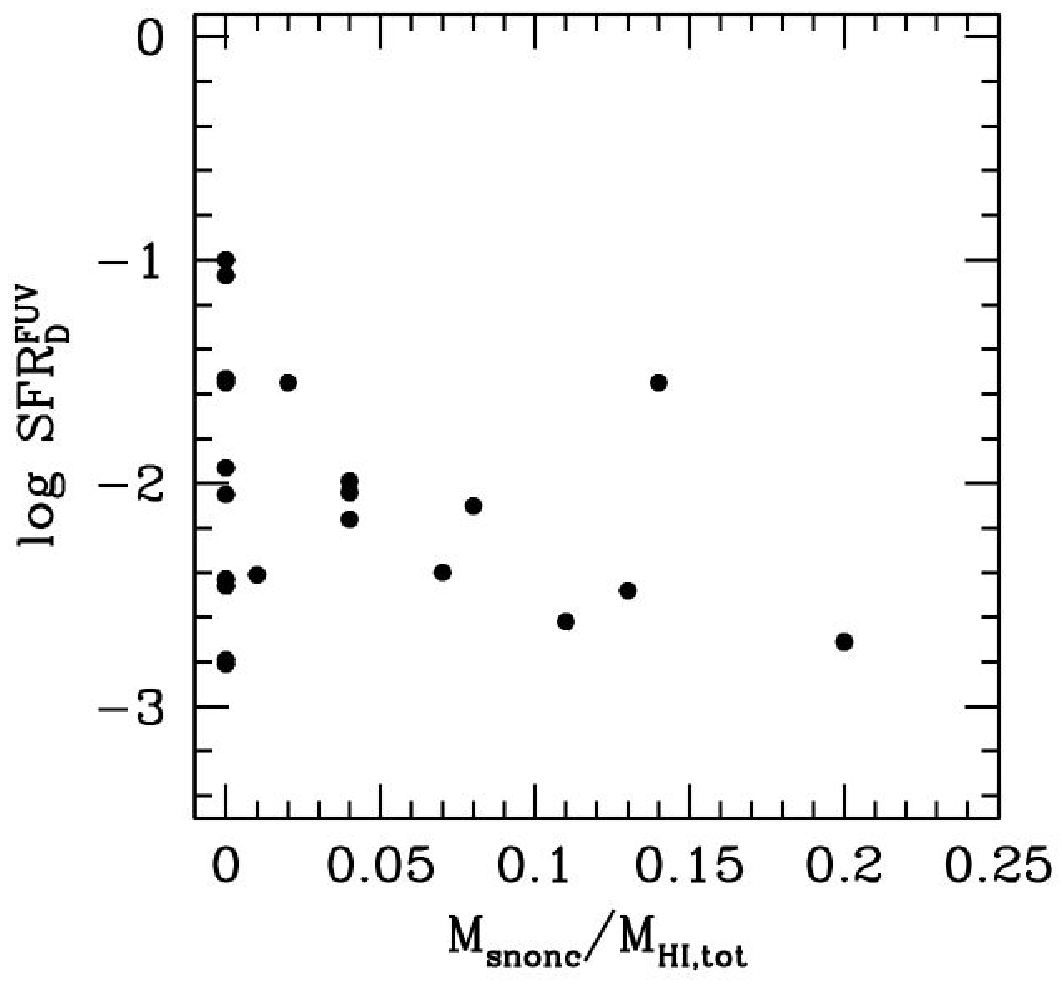}
\includegraphics{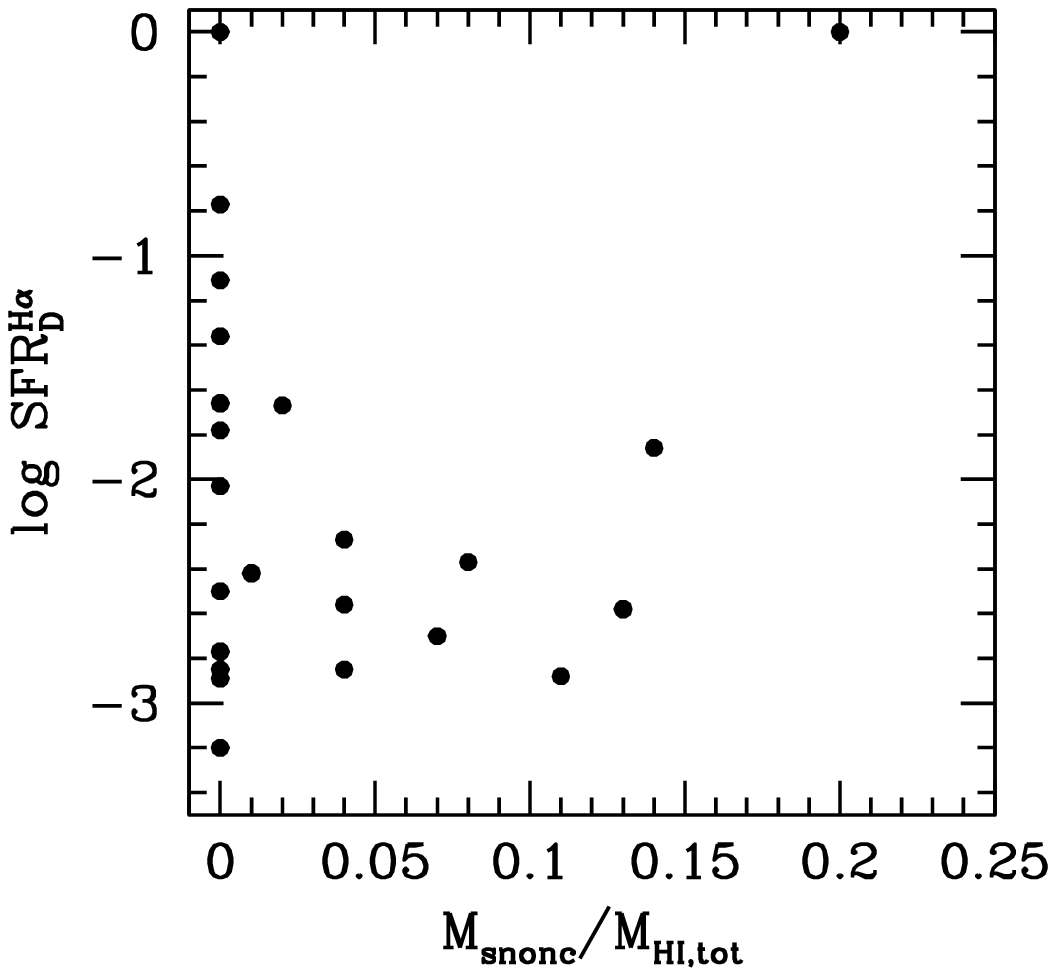}
\vskip 2.5truein
\caption{
Galactic star formation rate surface density, in units of $M$\solar\ yr$^{-1}$ kpc$^{-2}$
plotted against fraction of \HI\ gas engaged in noncircular motions
for each galaxy.
Left: The star formation rate was determined from {\it GALEX} FUV luminosity and is normalized to the area within one disk scale length,
from \citet{ludka}. % with distances given in \citet{lt12}.
Right: Same as the panel on the left but for a star formation rate determined from \ha\ emission, from \citet{he04}.
\label{fig-sfr}}
\end{figure}

We plot the star formation rate surface density
of each galaxy against the ratio of mass of \HI\ in that galaxy engaged in significant noncircular
motion to the total \HI\ mass of the galaxy. This is shown in Figure \ref{fig-sfr} with the
star formation rate surface density determined from FUV emission on the left and that from \ha\ on the right.
%There is not much of a correlation. 
Galaxies without gas engaged in non-circular motions (at $M_{\rm snonc}/M_{\rm HI,tot}=0$),
cover the full range of star formation rates (SFRs), but for the rest of the galaxies, there is a slight decrease in 
FUV SFR surface density with increasing $M_{snonc}/M_{HI,tot}$
although this is not evident for the \ha\ SFR surface density. The FUV trend is opposite to the expectation that
noncircular motions might facilitate star formation.
On the other hand, if the presence of large amounts of snonc gas results in starbursts that are short lived, we
might be less likely to catch those events.

On a local level, too, there is no pattern of correspondence between star forming regions and strong noncircular motion gas.
Generally speaking in a given galaxy, a few nonciruclar motion regions might be coincident with FUV or \ha\ emission, but other regions are not
and there can be lots of star formation in a galaxy not connected to gas moving differently from the bulk rotation.

% fig18

\subsection{Relationship of noncircular motion gas to holes}

\begin{figure}[t!]
\vskip -0.85truein
\includegraphics[scale=1.5]{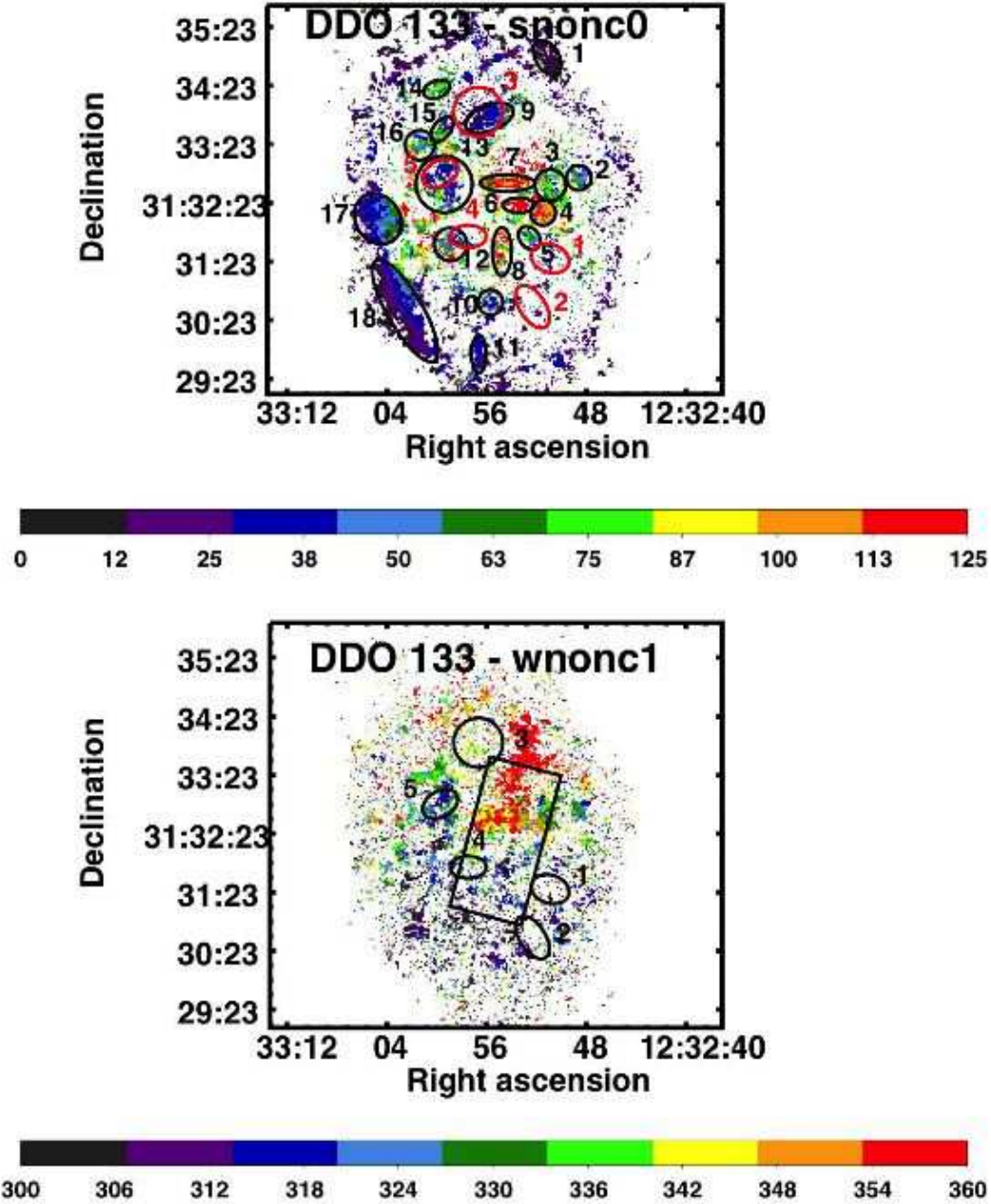}
\vskip -0.25truein
\caption{\HI\ holes catalogued by \citet{pokhrel16} shown on the snonc0 and wnonc1 images of DDO 133.
Top panel: The holes are shown in red and the snonc features are shown in black.
The units of the colorbar are Jy beam$^{-1}$ m s$^{-1}$.
Bottom panel: The holes are outlined in black and the stellar bar is outlined with a black rectangle.
The units of the colorbar are \kms.
\label{fig-holes}}
\end{figure}

Holes in the \HI\ gas of the LITTLE THINGS dwarfs have been catalogued by \citet{pokhrel16} \citep[see][for the methodology 
in identifying the holes]{bagetakos11}.
As an example, in Figure \ref{fig-holes} we plot the location of the 8 holes in DDO 133 and compare those to
the location of snonc and wnonc features. 
We see that three of the gas holes overlap with snonc features. 
The most interesting hole is number 3, which overlaps with snonc region 9 and is adjacent to the northeast 
corner of the stellar bar and to the east of the northern part of a wnonc feature at the end of the bar (large red area in bottom panel of Figure \ref{fig-holes}).
Expansion of the shell around hole 3 might have contributed to the snonc and wnonc features,
although currently the hole has blown out of the disk and is not expanding \citep{pokhrel16}.
Hole number 5 also appears to be connected with wnonc gas seen in the bottom panel of Figure \ref{fig-holes} and snonc region number 13,
and hole number 4 overlaps with snonc region 12.
%However, generally the \HI\ holes have little connection with snonc motion gas.

% fig19

\subsection{Relationship of noncircular motion gas to velocity dispersion}

In the bottom right panel of Figures \ref{fig-maps1} to \ref{fig-maps11} we plot the snonc features on
the \HI\ velocity dispersion (moment 2) maps of the galaxies.
Here we are looking for correlations between the snonc features with velocity dispersion 
that would indicate that these motions generate excess turbulence.
\citet{ashley14}, for example, found excess turbulence at the impact points of accreting gas in IC 10.
In most galaxies there are some snonc features in regions at the upper range of turbulence values in the galaxy.
For example, in CVnIdwA both regions are adjacent to the two regions of higher velocity dispersion (red in Figure \ref{fig-maps1}).
Other snonc regions are located in the central parts of the galaxies where the velocity dispersion is overall higher, as in DDO 47 (Figure \ref{fig-maps2}), 
and the association is not likely specific to the snonc nature of the regions;
the higher velocity dispersion regions are located throughout the galaxy and the snonc regions appear to
be randomly associated, as in DDO 50 (Figure \ref{fig-maps3});
or a few snonc regions are associated with higher velocity dispersion but many others are not, as in DDO 70 (Figure \ref{fig-maps5}).
Thus, generally there is no systematic correlation between snonc regions and regions of high velocity dispersion.

However, there are two exceptions of note.
In DDO 168 (Figure \ref{fig-maps8}), all of the snonc features are located on the eastern edge of 
a large part of the galaxy where the velocity dispersion exceeds 14 \kms. There is nothing 
obviously unusual in that region of the galaxy at other wavelengths, so it is not clear why this large section of the galaxy has such a high 
velocity dispersion.
In DDO 210 (Figure \ref{fig-maps9}), just outside the snonc regions to the northeast and southeast
are areas where the velocity dispersion of the gas is $>11$ \kms. These high velocity dispersion regions line the edges
of the large-scale arc structures.

\section{Summary} \label{sec-summary}

We have examined maps of the gas engaged in noncircular motions in 22 of the LITTLE THINGS
sample of nearby dIrr galaxies. The \HI\ data cubes have been 
deconvolved into \HI\ in ordered ``bulk'' motions,
\HI\ engaged in noncircular motions that has higher intensity peaks than the underlying bulk motion gas (``snonc''),
and \HI\ engaged in noncircular motions that has lower intensity peaks than the underlying bulk motion gas (``wnonc'').
Maps of these kinematic components consist of integrated moment 0, velocity field moment 1, and velocity dispersion 
moment 2 maps. We found significant regions of snonc motion gas in half of the galaxies.

In DDO 133 we found a ``crinkle'' pattern in the velocity field isophotes that is characteristic of 
streaming motions around the obvious stellar bar potential seen in the optical. 
In addition concentrations of gas engaged in noncircular motions and star-forming regions
are found at the northern end of the bar. 
The same velocity field pattern is found in DDO 47, but no stellar bar is obvious.

DDO 47 and DDO 133 also have large-scale filamentary structures found in 
the outer disks with strong noncircular motions. Neither filament is connected with any
star-forming regions. 
These filaments could be transient instabilities in the gas.
Given the timescales, the likely mechanism for destruction of the filament in DDO 133 is
dispersion due to the velocity dispersion of the gas.
The destruction mechanism in DDO 47 could be either dispersion or shearing perturbations.

We compared the location of noncircular motion gas features with the location 
of star-forming regions. We only see a correlation at the northern end of the bar in DDO 133, where
streaming motions of the gas around the bar has probably facilitated cloud formation. 
There is also no correlation between the integrated galactic
star formation rate surface density and the ratio of gas mass engaged in noncircular motion to total \HI\ mass of the galaxy or
on a local scale between holes in the gas and snonc gas.
In most cases the snonc gas is not spatially related to an enhancement in the gas velocity dispersion.
However, there is an unusual correlation of snonc gas in DDO 168 and a large portion of the galaxy with higher 
velocity dispersion in the gas than elsewhere in that galaxy.

\acknowledgments

LL appreciates funding from the National Science Foundation grant
AST-1461200 to Northern Arizona University for Research Experiences for Undergraduates
summer internships and Dr.\ David Trilling for running the NAU REU program in 2018.
We wish to thank Dr.\ Michael West for discussions in the early stages of this project.
Obtaining and reducing the LITTLE THINGS data that were used here were funded in part by the National Science Foundation through 
grants AST-0707563, AST-0707426, AST-0707468, and AST-0707835 to US-based LITTLE THINGS team members 
and with generous technical and logistical support from the National Radio Astronomy Observatory.

Facilities: \facility{VLA}

\end{document}